\def\xflux{~erg~cm$^{-2}$~s$^{-1}$}
\def\xlum{~erg~s$^{-1}$}

\documentclass[preprint,longnamesfirst]{aastex}
\usepackage{natbib}
\usepackage{emulateapj5,apjfonts}
\tighten

\journalinfo{The Astronomical Journal, 124: ??1--?12, 2002 November, astro-ph/0207433}
\slugcomment{Received 2002 May 13; accepted 2002 July 19}
\shortauthors{BAUER ET AL.}
\shorttitle{X-RAY AND RADIO SOURCE POPULATIONS IN THE CDF-N}

\begin{document}

\title{The {\it Chandra} Deep Field North Survey. XII. The Link Between 
Faint X-ray and Radio Source Populations}

\author{
F.~E.~Bauer,\altaffilmark{1}
D.~M.~Alexander,\altaffilmark{1}
W.~N.~Brandt,\altaffilmark{1}
A.~E.~Hornschemeier,\altaffilmark{1}
C.~Vignali,\altaffilmark{1}
G.~P.~Garmire,\altaffilmark{1}
and D.~P.~Schneider\altaffilmark{1}
}

\altaffiltext{1}{Department of Astronomy \& Astrophysics, 525 Davey Lab, 
The Pennsylvania State University, University Park, PA 16802.}

\begin{abstract}
We investigate the relationship between faint X-ray and 1.4~GHz radio
source populations detected within 3$\arcmin$ of the Hubble Deep Field
North using the 1~Ms {\it Chandra} and 40~$\mu$Jy VLA surveys. Within
this region, we find that $\approx$42\% of the 62 X-ray sources have
radio counterparts and $\approx$71\% of the 28 radio sources have
X-ray counterparts; thus a 40~$\mu$Jy VLA survey at 1.4~GHz appears to
be well-matched to a 1~Ms {\it Chandra} observation. Among the
different source populations sampled, we find that the majority of the
18 X-ray detected emission-line galaxies (ELGs) have radio and
mid-infrared {\it ISOCAM} counterparts and appear to be luminous
star-forming galaxies at $z=0.3$--1.3. Importantly, the radio-detected
ELGs make up $\approx$35\% of the X-ray source population at
0.5--8.0~keV X-ray fluxes between $\approx(1$--$5)\times10^{-16}$
\xflux\ and signal the emergence of the luminous, high-$z$ starburst
galaxy population in the X-ray band. We find that the
locally-determined correlation between X-ray luminosities and 1.4~GHz
radio luminosity densities of the late-type galaxies can easily be
extended to include the luminous intermediate-redshift ELGs,
suggesting that the X-ray and radio emission processes are generally
associated in star-forming galaxies. This result implies that the
X-ray emission can be used as an indicator of star formation rate for
star-forming galaxies. Finally, we show that there appear to be two
statistically distinct types of {\it ISOCAM}-detected star-forming
galaxies: those with detectable radio and X-ray emission and those
without. The latter type may have stronger mid-infrared emission-line
features that increase their detectability at mid-infrared
wavelengths.
\end{abstract}


\section{Introduction}\label{intro}

At bright X-ray fluxes ($\ga10^{-15}$ \xflux), almost all
extragalactic X-ray point sources are found to be active galactic
nuclei \citep[AGN; e.g.,][]{Bade1998, Schmidt1998, Akiyama2000,
Bauer2000, Lehmann2001}. At the faintest currently achievable levels,
however, a large population of ``normal'' galaxies emerge
\citetext{\citealp[e.g.,][]{Brandt2001a}, hereafter Paper~IV;
\citealp{Hornschemeier2001}, hereafter Paper~II; \citealp{Tozzi2001}}. 
These X-ray faint ``normal'' galaxies are often optically bright
($I\la22$) and are predominantly identified as narrow emission-line
galaxies with redshifts of $z=0.3$--$1.3$. The majority appear to have
faint 15$\mu$m {\it ISOCAM} counterparts \citep[][hereafter Paper
XI]{Alexander2002}, suggesting a close association with the strongly
evolving luminous infrared (IR) starburst galaxy population
\citep[e.g.,][]{Elbaz2002}. X-ray/IR correlations have also been found
locally for late-type normal spiral and irregular galaxies
\citep[e.g.,][]{Fabbiano1988, David1992, Shapley2001}. Thus the
study of optically bright, X-ray faint galaxies may provide important
clues about the nature of star formation on cosmic timescales.

An equally good correspondence between X-ray faint galaxies and faint
radio sources is perhaps to be expected, given (1) the X-ray/radio
correlation found for local late-type normal spiral and irregular
galaxies \citep[e.g.,][]{Fabbiano1988, Shapley2001, Ranalli2002}, and
(2) the tight relationship between radio and mid-IR/far-IR (MIR/FIR)
emission found for star-forming galaxies both locally and at moderate
redshifts \citep[e.g.,][]{Helou1985, Condon1991, Price1992,
Shapley2001, Elbaz2002, Garrett2002}. Such a correlation can be
understood physically, since X-ray and radio emission are both tied to
the evolution of massive stars in the form of mass-transfer in
binaries and supernovae, respectively \citep[e.g.,][]{Petre1993,
Bressan2002}. Indeed, within the Hubble Deep Field North itself
\citep[\hbox{HDF-N};][]{Williams1996}, the faint radio and X-ray
populations share a large overlap ($\sim$70\%; Paper~IV).

In this paper, we investigate in detail the observational properties
of the faint X-ray/radio matched sources in the vicinity of the \hbox{HDF-N}
using two of the deepest surveys ever performed at their respective
wavelengths, the 1~Ms {\it Chandra} Deep Field North X-ray
\citep[\hbox{CDF-N};][hereafter Paper~V]{Brandt2001b} and the 40~$\mu$Jy VLA
1.4~GHz radio \citep[][hereafter R00]{Richards2000} surveys. This work
extends the results of Paper~IV to a larger area and to deeper X-ray
fluxes, and it examines in more detail the relationships between the
X-ray, radio, and IR emission. To facilitate comparison with previous
X-ray/IR matching results in Paper~XI, we adhere to much of the
procedure outlined in that study. We describe the selection of our
X-ray and radio samples in $\S$\ref{sample}, compare and contrast
these samples in $\S$\ref{properties} and $\S$\ref{categories}, and
discuss implications in $\S$\ref{discussion}. Throughout this paper,
we adopt $H_{0}=65$~km~s$^{-1}$~Mpc$^{-1}$, $\Omega_{\rm
M}=\onethird$, and $\Omega_{\Lambda}=\twothirds$. The Galactic
column~density toward the
\hbox{CDF-N} is $(1.6\pm0.4)\times10^{20}$ cm$^{-2}$ \citep{Stark1992}. 
Coordinates are for the J2000 epoch.

\section{Sample Construction}\label{sample}

To compare X-ray and radio sources in the \hbox{CDF-N}, we restricted
our focus to a $3\arcmin$ radius region (28.3~arcmin$^2$) centered on
the \hbox{HDF-N} ($\alpha_{2000} = 12^{\rm h} 36^{\rm m} 49\fs4$,
$\delta_{2000} = +62^{\circ} 12\arcmin 58\arcsec$). This area was
chosen because it has the deepest and most uniform coverage at both
X-ray and radio wavelengths, and it contains a wealth of follow-up
resources including hundreds of spectroscopic redshifts
\citep[e.g.,][]{Cohen2000, Dawson2001} and deep imaging in most
astronomical wavelength bands \citep[for a review
see][]{Ferguson2000}. These ancillary data are extremely important for
assessing the nature of the X-ray and radio source populations. Of
particular relevance, much of the $3\arcmin$ radius region overlaps
with the {\it ISOCAM} 15$\mu$m observations \citep{Aussel1999}.

\subsection{X-ray-Detected Sample}\label{xsample}

The X-ray selected sample was chosen from sources detected in the 1~Ms
\hbox{CDF-N} catalog of Paper~V in one or more of the 0.5--2.0~keV
(soft), 2--8~keV (hard), and 0.5--8.0~keV (full) bands using {\sc
  wavdetect} \citep{Freeman2002} with a false-positive probability
threshold of $1\times10^{-7}$ (for details see Paper~V). Within
$3\arcmin$ of the \hbox{HDF-N}, there are 62 X-ray sources detected
down to an on-axis point-source full-band flux limit of $\approx
1\times 10^{-16}$ \xflux.  Table~\ref{tab:xray-sources} lists the
properties of the sources in the X-ray-detected sample.

\subsection{Radio-Detected Sample}\label{rsample}

The radio-selected sample was chosen from sources detected at 1.4~GHz
in the VLA catalog of R00. Within $3\arcmin$ of the \hbox{HDF-N},
there are 28 radio sources detected down to a 5$\sigma$ flux density
limit of $\approx$~40~$\mu$Jy. This region also has deep VLA
observations at 8.5~GHz \citep[][hereafter R98; R00]{Richards1998}. 
The 8.5~GHz observations probe down to a 5$\sigma$ flux density limit
of $\approx$~9--20~$\mu$Jy, allowing spectral indices to be calculated
for the majority of radio sources. We note that the sensitivity of the
1.4~GHz observations over the region of interest is very uniform,
while that of the 8.5~GHz sharply falls off past a radius of
$\approx$~2$\arcmin$. However, because the 1.4~GHz observations were
performed using the VLA A-array, the completeness limit for extended
radio sources larger than $\sim 3.5\arcsec$ is somewhat higher ($\la
50$~$\mu$Jy; see section 3.1 of R00 for details). Thus, the R00
catalog may be incomplete for galaxies closer than $z\sim0.3$. 
Table~\ref{tab:radio-sources} gives the properties of the sources in
the radio-detected sample.

\subsection{Source Matching}\label{sample_match}

Source matching was performed by cross-identifying X-ray and radio
sources to $I$-band counterparts detected by 
\citeauthor{Barger2002}~\citetext{\citeyear{Barger2002}; for X-ray sources
reported in Paper~V} or
\citeauthor{Bauer2002}~\citetext{\citeyear{Bauer2002}; for radio
sources which lack 1~Ms catalog counterparts}. Radio and X-ray sources
were matched to optical sources using a $1\arcsec$ search radius,
except in three cases where the extent of the optical galaxy was quite
large: CXOHDFN~J123637.0$+$621134 ($\approx 10\arcsec$ optical diameter,
1\farcs3 X-ray/radio offset) and CXOHDFN~J123641.8$+$621132
($\approx 5\arcsec$ optical diameter, 1\farcs1 X-ray/radio offset). To within
positional errors, however, the X-ray and 1.4~GHz radio emission from
the matched sources were spatially coincident in all
cases.\footnote{In two galaxies, CXOHDFN~J123641.8$+$621132 and
CXOHDFN~J123648.3$+$621426, the X-ray source is coincident with lower
significance 1.4~GHz detections but not with higher significance
8.5~GHz detections (offsets are $\approx2\arcsec$ but still within the
extents of the galaxies).} Note that the 1$\sigma$ positional
uncertainties between the three wavelength bands are $\la0\farcs6$
within the $3\arcmin$ radius region.

There are 18 sources in common between the X-ray- and radio-detected
samples. Given the relatively low surface density of X-ray and radio
sources, however, we can search for lower significance radio and X-ray
sources within the respective samples without introducing a
significant number of spurious detections.

Lower significance radio counterparts to X-ray sources were determined
in two ways. The first and most direct method was to measure
3--5$\sigma$ radio emission at the positions of unmatched X-ray
sources using the {\sc aips} task {\em jmfit} on the publicly
available 1.4~GHz radio image (R00).\footnote{This image is available
at http://www.cv.nrao.edu/$\sim$jkempner/vla-hdf/.} Unfortunately,
this image does not cover the entire $3\arcmin$ region, and 18 sources
from the X-ray-detected sample therefore have no measurable 1.4~GHz
flux density limits. Nonetheless, five additional 1.4~GHz radio
counterparts were found (see Table~\ref{tab:radio-sources} for
details). The probability of finding a $>3\sigma$ 1.4~GHz radio source
at the position of an unmatched X-ray source is $<0.02$. The second
method was to match X-ray sources to radio sources detected at 8.5~GHz
(R98) but not at 1.4~GHz. Three additional radio sources were found in
this manner, and 1.4~GHz flux density upper limits were assigned to
these sources assuming a conservative spectral index of $\alpha=0.8$
\citep[the typical value for a starburst galaxy;
e.g.,][]{Yun2001}.\footnote{The radio spectral index is defined as
$S_{\nu} \propto \nu ^{-\alpha}$. Lower values of $\alpha$ thus yield
tighter constraints on the 1.4~GHz flux density.} One of these
sources, CXOHDFN~J123637.0$+$621134, has a 99~$\mu$Jy counterpart in
the WSRT 1.4~GHz map of \citet{Garrett2000}, and this WSRT value has
been used in lieu of a VLA 1.4~GHz upper limit; see
Table~\ref{tab:xray-sources}.\footnote{This source has an optical
extent of $\sim6$--$10\arcsec$ and was perhaps ``resolved out'' by the
VLA 1.4~GHz observations.} Including these additional sources, there
are a total of 26 ($42^{+10}_{-8}$\%) X-ray sources with radio
counterparts.\footnote{All errors are taken from Tables 1 and 2 of
\citet{Gehrels1986} and correspond to the $1\sigma$ level; these were
calculated assuming Poisson statistics.}

Lower significance X-ray counterparts to radio sources were determined
using {\sc wavdetect} on the soft-, hard-, and full-band 1~Ms
\hbox{CDF-N} X-ray images with a false-positive probability threshold
of $1\times10^{-5}$. Two additional low-significance X-ray matches to
radio sources were found. The probability of finding an X-ray source
at the position of a known radio source and with the above {\sc
  wavdetect} threshold at random is $<0.003$. Including these
additional sources, there are a total of 20 ($71^{+20}_{-16}$\%) radio
sources with X-ray counterparts. This matching fraction is
significantly higher than was reported with 221.9~ks (Paper~II) but
comparable to that found within the smaller \hbox{HDF-N} region with
479.7~ks (Paper~IV).

\section{Basic Properties of X-ray/Radio Matched Objects}\label{properties}

The overlap between the X-ray- and radio-detected samples is shown in
Figure~\ref{fig:x_vs_r}, with the dashed lines indicating constant
ratios of X-ray-to-radio flux, $f_{\rm X}/f_{\rm 1.4~GHz}$.\footnote{A
44~MHz bandwidth was used to calculate the radio flux (R00), which in
turn was used to calculate the flux ratio, $f_{\rm X}/f_{\rm
1.4~GHz}$.} We characterize the sources in terms of their full-band
X-ray fluxes, rather than their soft- or hard-band fluxes as has
traditionally been done in the past, because (1) the vast majority of
the sources in the X-ray sample are detected in the full band (94\%
versus 81\% in the soft band and 66\% in the hard band) and (2) the
full band more effectively separates obscured AGN from ``normal''
galaxies, especially when combined with optical magnitude information,
since it incorporates both soft- and hard-band X-ray fluxes. For
instance, a galaxy and a moderately-obscured AGN could potentially
have comparable soft-band X-ray fluxes, but because 
\vspace{-0.1in}
\centerline{
\includegraphics[width=9.0cm]{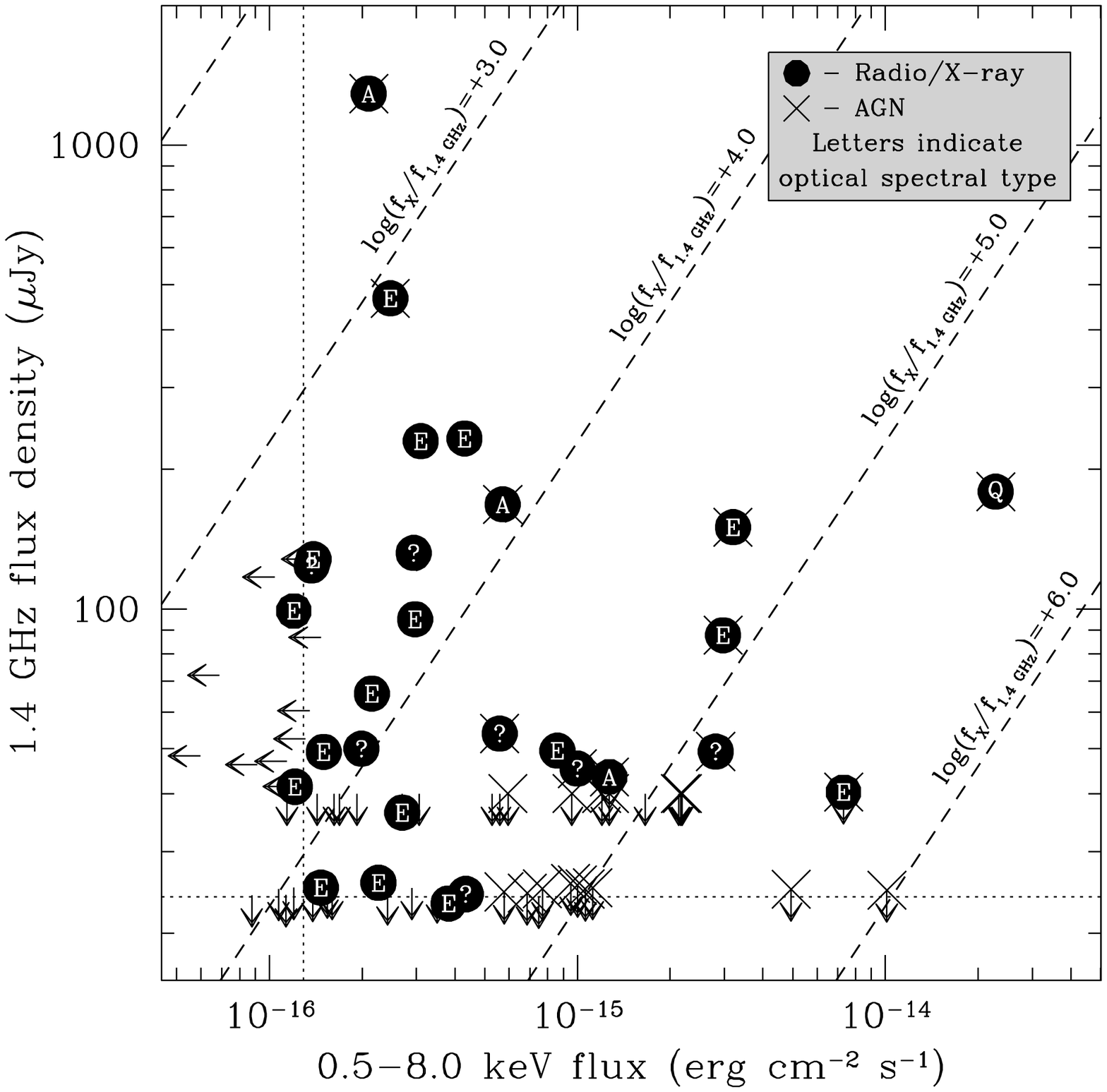}
}
\vspace{-0.2cm} 
\figcaption[fig1.ps]{
A comparison of 1.4~GHz flux density versus full-band X-ray flux for
all X-ray and radio detected sources (i.e., detections at either
1.4~GHz or 8.5~GHz). Filled circles indicate X-ray/radio matched
sources, while arrows signify source upper limits ($3\sigma$ or
$5\sigma$ depending on if the source lies in within the publicly
available 1.4~GHz image). Crosses indicate that AGN activity dominates
either the X-ray or radio band (see $\S$\ref{properties}). Letters
represent spectroscopic identifications
\citep[e.g.,][]{Cohen2000, Dawson2001}, with ``E's'', ``A's'', and
``Q's'' denoting ELGs, ALGs, and broad-line AGN, respectively, while
``?'s'' designate sources which remain unidentified (see
Tables~\ref{tab:xray-sources} and \ref{tab:radio-sources}); for
clarity, however, letters are plotted only for X-ray/radio matched
sources. Lines of constant flux ratio are denoted by diagonal dashed
lines, while the approximate detection limits of the X-ray and radio
surveys are shown by vertical and horizontal dotted lines,
respectively.
\label{fig:x_vs_r}}
\vspace{0.2cm} 

\noindent of their different
spectral slopes (typically $\Gamma\sim2.0$ versus 1.0, respectively),
the obscured AGN will have a much larger full-band flux and be easily
identified.

We have adopted the source classification scheme presented in Paper~XI
with the exception that we further use the radio properties of the
sources to identify AGN activity. This scheme is pivotal to all that
follows and is thus repeated here. Since nearly all of the sources
with $I<23$ have spectroscopic redshifts and classifications
\citep[e.g.,][]{Cohen2000, Dawson2001}, sources were classified into
three loose spectral types: emission-line galaxies (ELGs), and
absorption-line galaxies (ALGs), and AGN-dominated sources. These
categories are intended to identify qualitatively the bulk of the
optical emission from a particular galaxy as either from active star
formation, passively evolving stars, or AGN activity. Note, however,
that in classifying the sources as ELGs and ALGs, we have ignored some
of the subtleties of the optical spectral types --- such as whether
the emission-line spectra have \ion{H}{2}-like emission lines
indicative of active star formation or exhibit only a weak \ion{O}{2}
$\lambda$3727 doublet --- since the spectra remain largely
unpublished. Thus we caution that spectral misclassifications are
possible for some sources.

We classified sources as ELGs if they have optical spectral
classifications of ``E'', ``I'', ``EI'' or ``EA'' in
\citet{Cohen2000}. Likewise, we classified sources as ALGs if they
have a classification of ``A'' in \citet{Cohen2000}. AGN-dominated
sources were identified based on one of the following criteria: a
spectroscopic identification as a broad-line AGN \citep[i.e., a ``Q''
spectral type from][]{Cohen2000}, a rest-frame 0.5--8.0~keV luminosity
$>3\times10^{42}$~\xlum, an effective X-ray photon index $\Gamma<1.0$
(indicative of obscured AGN), or radio properties indicative of AGN
\citetext{i.e., spectral indices $\alpha<0.4$--$0.5$, radio
jets/lobes, or variability; R98; \citealp{Richards1999a}, chap. 4,
hereafter R99; R00}. Any ELGs or ALGs that were determined to contain
AGN were classed as AGN-dominated. Tables~\ref{tab:xray-sources} and
\ref{tab:radio-sources} list the spectral types of each source. In
Figure~\ref{fig:x_vs_r}, optical spectral types are represented by the
letters ``E'' for ELGs, ``A'' for ALGs, ``Q'' for broad-line AGN, and
``?'' if the spectral type remains unknown, while AGN-dominated sources are
indicated by large crosses.

Since the ratio of X-ray-to-radio flux found for classical AGN can
span {\it several} orders of magnitude \citep[e.g.,][Fig. 
4.10]{Kellermann1989, Brinkmann2000, Bauer2001}, it is not surprising
to see such a large spread among the AGN in Figure~\ref{fig:x_vs_r}
[$\log (f_{\rm X}/f_{\rm 1.4~GHz})\sim3$--6]. Less stringent AGN
activity constraints can be placed on the fainter X-ray and radio
sources because of their low signal-to-noise and some are therefore
likely to host low-luminosity AGN. We note, however, that the large
number of X-ray/radio matched ELGs at faint X-ray fluxes and radio
flux densities is consistent with the emergence of a ``normal,''
star-forming galaxy population at moderate redshifts
\citetext{\citealp[e.g.,][]{Windhorst1985}; R98; Paper~II; \citealp{Tozzi2001}}.
Interestingly, there appears to be a steep rise in the number of radio
matches to X-ray sources near the limit of the 1.4~GHz survey,
suggesting that slightly deeper radio observations may detect several
of the remaining X-ray sources.

The ubiquity of star formation within these X-ray/radio matched
sources is most clearly demonstrated by comparing 
\vspace{-0.1in}
\centerline{
\includegraphics[width=9.0cm]{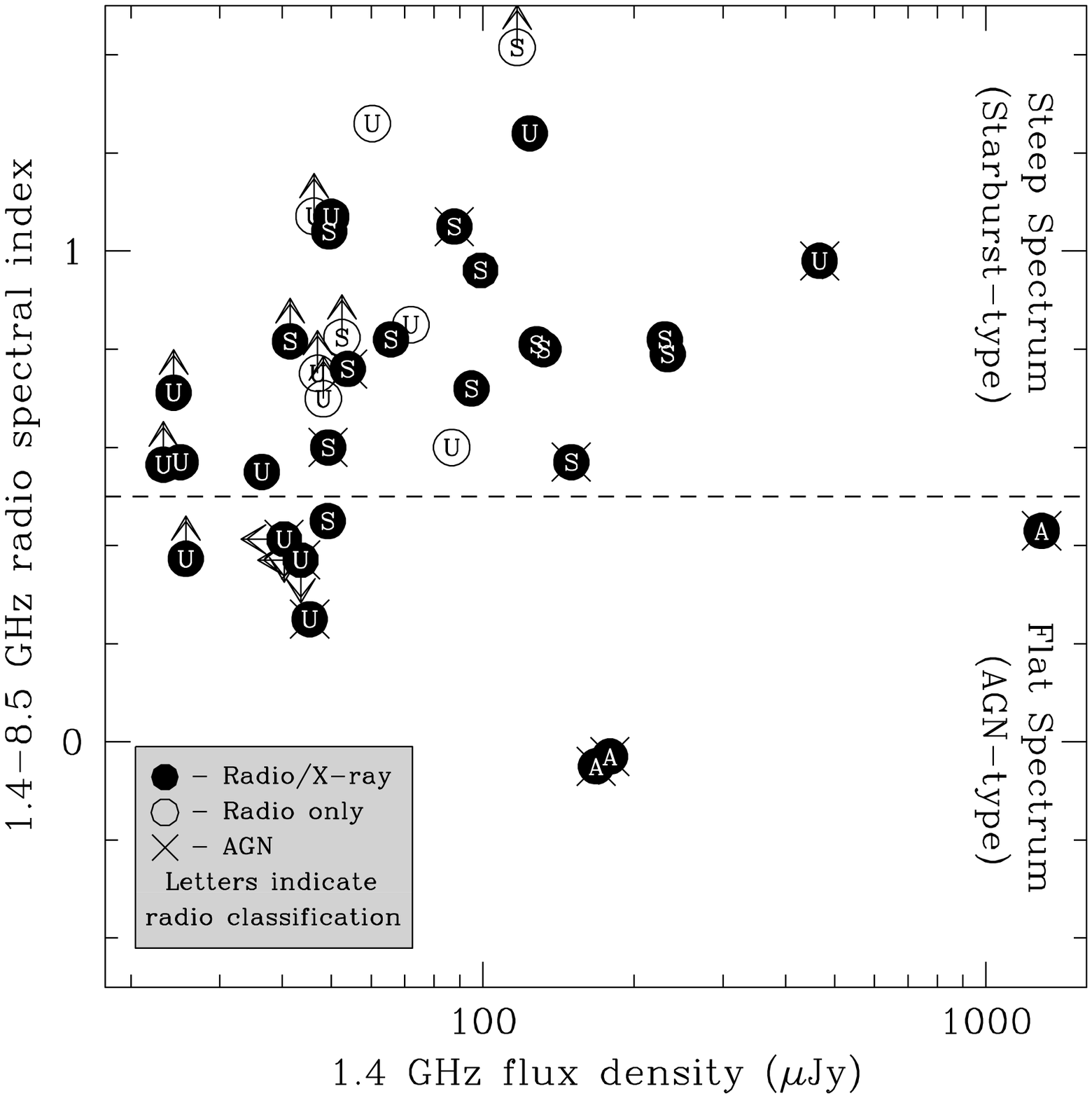}
}
\vspace{-0.2cm} 
\figcaption[fig2.ps]{
A comparison of 1.4--8.5~GHz spectral index ($\alpha$) versus
1.4~GHz flux density for all radio sources. Filled circles indicate
X-ray/radio matched sources, open circles denote unmatched radio
sources, and arrows signify approximate $3\sigma$ source upper or
lower limits. Crosses indicate that AGN activity dominates either
the X-ray or radio band (see $\S$\ref{properties}). Letters
represent the classifications of R99: ``S's'' and ``A's'' denote
starbursts and AGNs, respectively, while sources with ``U's'' remain
unidentified. The horizontal dashed line denotes the typical
spectral index, $\alpha=0.5$, used to separate steep spectrum
(starburst-type) and flat spectrum (AGN-type) systems (e.g., R00).
The two X-ray/radio matches detected only at 8.5~GHz are shown as
upper limits in both axes.
\label{fig:r_vs_ar}}
\vspace{0.2cm} 

\noindent the 1.4--8.5~GHz radio spectral index to the 1.4~GHz flux 
density, as shown in Figure~\ref{fig:r_vs_ar}. Plotted are radio
sources detected at either 1.4 or 8.5~GHz in the X-ray and radio
samples, along with the AGN classifications outlined above ({\it
crosses}) and the radio classifications from R99 ({\it letters}). 
These radio classifications --- starburst-type sources (S), AGN
candidates (A), and sources that could not be easily classified
(referred to as unknown, or U) --- are based primarily on radio
spectral indices, morphologies, and angular sizes, but they also take
into account optical and IR properties in some cases \citetext{see
\citealp{Muxlow1999} and R99}. The radio classifications are intended
to identify qualitatively the origin of the bulk of the radio emission
from a particular galaxy and do not preclude the possibility of
starburst-type sources having embedded AGN or AGN-type sources having
stellar activity. When possible we have used the matched radio
spectral indices provided in Table~1 of R99, where the 1.4~GHz and
8.5~GHz flux densities used to calculate the spectral indices were
measured in their respective 3\farcs5 convolved images. For the
remaining sources --- i.e., those with 3$\sigma$ detections at either
1.4~GHz or 8.5~GHz --- we have estimated the spectral indices using
the publicly available radio images; these lower significance
estimates are noted in Tables~\ref{tab:xray-sources} and
\ref{tab:radio-sources} and should be treated as approximate since the
images used to measure the fluxes were not matched in resolution. We
also caution that variability among the radio AGN is a legitimate
concern and could lead to inaccurate spectral index estimates since
the two radio bands were observed two years apart. A horizontal dashed
line denotes the typical spectral index, $\alpha=0.5$, separating
steep and flat spectrum radio sources. Steep spectral indices
($\alpha\ga0.5$) often indicate radio emission from star formation,
while flat ones ($\alpha\la0.5$) often indicate emission from
core-dominated AGN \citep[e.g.,][]{Kellermann1988}. Within our sample
area, there are 16 starburst-type, 3 AGN-type, and 17 unknown-type
sources. As expected, nearly all of the X-ray-detected non-AGN sources
are classified as starburst-type sources at radio wavelengths. 
Conversely, only approximately half of the X-ray/radio matched AGN are
classified as AGN-type sources at radio wavelengths, suggesting that
star formation activity is important in many of these AGN.

To explore further the nature of the X-ray and radio-selected samples,
we compare the distribution of full-band X-ray flux to $I$-band
magnitude (see Figure~\ref{fig:x_vs_i-radio}). Because of the
empirical relationship between X-ray and optical emission in AGN
\citep[e.g.,][]{Elvis1994}, this diagram highlights the degree and
nature of any AGN activity. In the X-ray-optical plane, classical AGN
typically have flux ratios of $-1 \la \log (f_{\rm X}/f_{\rm I}) \la
+1$, while star-forming galaxies and low-luminosity AGN tend to have
$\log (f_{\rm X}/f_{\rm I})\la-1$ \citetext{e.g.,
\citealp{Maccacaro1988}; \citealp{Stocke1991}; \citealp{Schmidt1998};
\citealp{Akiyama2000}; Paper~II; \citealp{Shapley2001}}. These regions
have been shaded and labeled for emphasis. We caution that although
star formation generally appears to be the dominant emission mechanism
for X-ray sources with $\log (f_{\rm X}/f_{\rm I})\la-1$ (e.g.,
Paper~II, Paper~XI), X-ray emission from low-luminosity or heavily
obscured AGN cannot be individually ruled out.

At first glance, the connection between X-ray and radio sources
appears to be rather complex, as X-ray/radio matched counterparts
(filled circles) span $\sim2$--3 decades in both X-ray flux and
optical magnitude. However, taking cues from the X-ray/IR matched
sources of Paper~XI, we notice a few interesting trends. 
First, the majority of the X-ray/radio matched sources that fall
within the ``galaxies'' region of Figure~\ref{fig:x_vs_i-radio} 
\vspace{-0.1in}
\centerline{
\includegraphics[width=9.0cm]{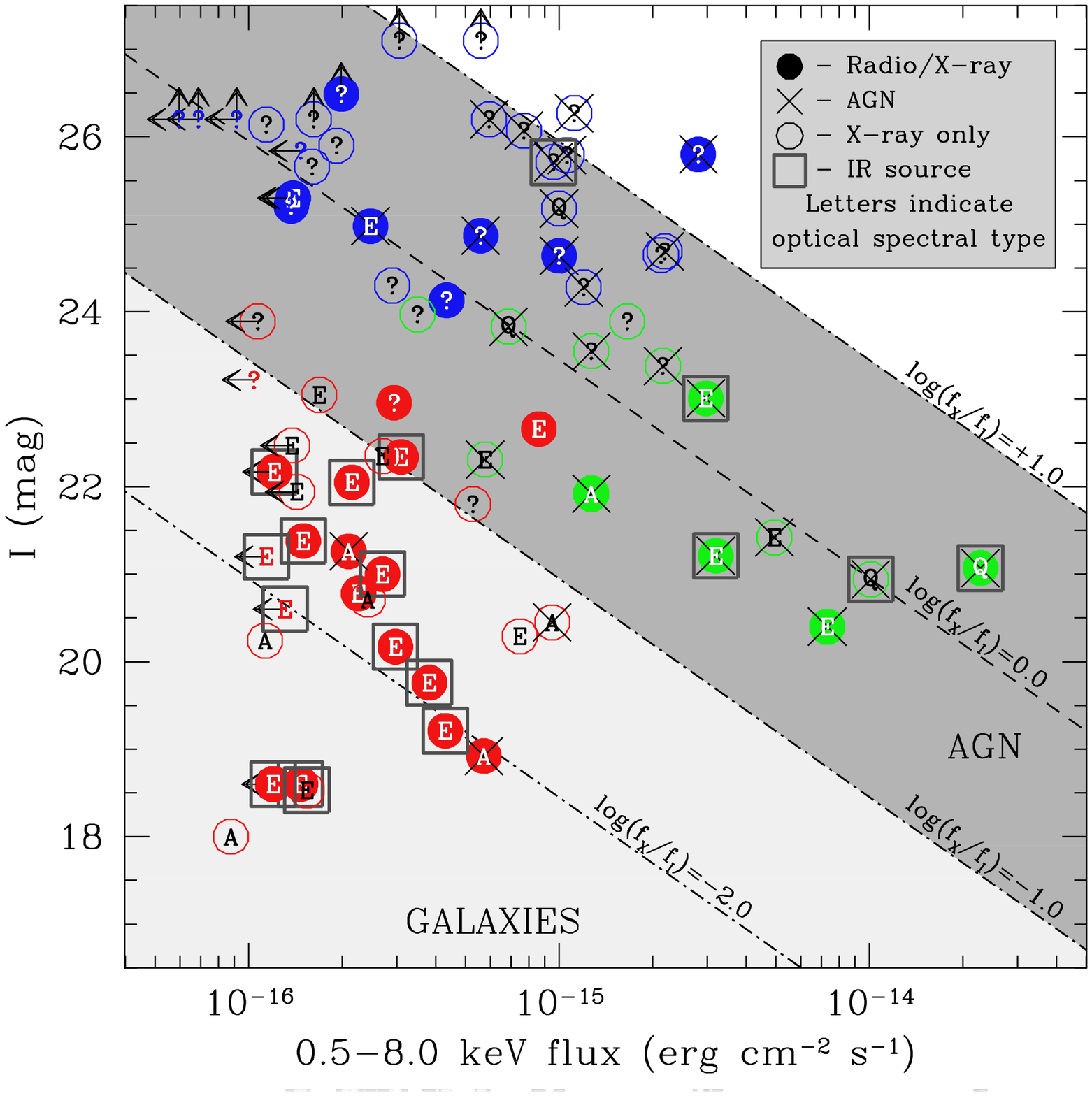}
}
\vspace{-0.0cm} 
\figcaption[fig3.ps]{
A comparison of $I$-band magnitude versus full-band X-ray flux for
X-ray- and radio-detected sources. Filled circles indicate X-ray/radio
matched sources, open circles denote unmatched X-ray sources, and
arrows signify $3\sigma$ source upper limits. The X-ray upper limits include
both sources which remain undetected at X-ray energies and sources
which are detected only in the soft or hard X-ray bands. As in
Figure~\ref{fig:x_vs_r}, crosses indicate that AGN activity dominates
either the X-ray or radio band, and sources with letters represent
spectroscopically identified sources (see $\S$\ref{properties}). 
Additionally, large concentric squares indicate sources with {\it
ISOCAM} 15$\mu$m counterparts (Paper~XI). Most of our $3\arcmin$
radius region overlaps with the {\it ISOCAM} 15$\mu$m observations;
the 29 sources outside the {\it ISOCAM} region are noted in
Tables~\ref{tab:xray-sources} and \ref{tab:radio-sources}. The colors
of the sources indicate whether they were classified as galaxies
(red), AGN (green), or optically faint sources (blue); see
$\S$\ref{properties}, as well as column~11 of
Table~\ref{tab:xray-sources} and column~13 of
Table~\ref{tab:radio-sources}.
\label{fig:x_vs_i-radio}}
\vspace{0.6cm} 

\noindent 
[i.e., $\log (f_{\rm X}/f_{\rm I})\la-1$] have IR counterparts (large
concentric squares) and are spectroscopically identified as ELGs,
compared to relatively few in the AGN region. As such, these sources
appear to be distinct from the typical X-ray emitting AGN. The IR
emission and ELG spectral types suggest that the X-ray emission in
these sources is likely to be associated with active star-forming
processes (Paper~XI).
Second, while there is no obvious division within the AGN population
[i.e., $\log (f_{\rm X}/f_{\rm I})\ga-1$], an observational division
occurs around $I\approx24$. The optically bright X-ray/radio matches
tend to be spectroscopically identified as galaxies or broad-line AGN
with $z\la1.0$, while the optically fainter ones remain largely
unclassified and perhaps comprise several different classes of object
\citep[see][hereafter Paper~VI]{Alexander2001}. 
Thus we divide the X-ray/radio matches into three basic categories:
galaxies, optically bright ($I<24$) AGN, and optically faint ($I\ge24$) sources
(see Figure~\ref{fig:x_vs_i-radio}, as well as column~12 of
Table~\ref{tab:xray-sources} and column~13 of
Table~\ref{tab:radio-sources}). The three source types are highlighted
in Figure~\ref{fig:x_vs_i-radio} by the colors red, green, and blue,
respectively. 

\section{Categories of X-ray/Radio Matched Objects}\label{categories}

Below we provide constraints on the basic nature of the sources in the
``galaxies'', ``optically bright AGN'', and ``optically faint
sources'' categories.

\clearpage
\subsection{Galaxies}\label{galaxies}

The ``galaxies'' sample includes 21 X-ray sources that lie within the
$\log (f_{\rm X}/f_{\rm I})\la-1$ region of
Figure~\ref{fig:x_vs_i-radio} and a further six sources that lie
slightly above $\log (f_{\rm X}/f_{\rm I})=-1$; the latter sources are
included since they share many characteristics with the former (these
six sources are noted in Table~\ref{tab:xray-sources}). All but two of
the 27 objects have X-ray luminosities below $3\times10^{42}$~\xlum. 
Fifteen of the 27 (56\%) sources in the X-ray sample have radio
counterparts, while 15 of the 18 (83\%) sources in the radio sample
have X-ray counterparts. Thus, it appears that the 40~$\mu$Jy VLA
survey at 1.4~GHz is extremely well-matched to the 1~Ms {\it Chandra}
observation for the detection of galaxies.

In the nearby Universe, connections between X-ray, IR, and radio
emission appear to split along evolutionary lines, with the strongest
correlations arising from galaxies with active star formation, such as
late-type spirals and irregulars \citep[e.g.,][]{Shapley2001}. A
similar trend emerges in the \hbox{CDF-N} at moderate redshifts
($z=0.4$--1.3) for luminous IR starbursts and X-ray faint galaxies, as
evidenced by the findings of Paper~XI. X-ray faint galaxies with IR
counterparts are predominantly identified spectroscopically as ELGs
and exhibit no apparent X-ray/radio AGN activity, while those without
IR counterparts are primarily ALGs and exhibit low-to-moderate level
X-ray/radio AGN activity in roughly half of the cases. Since the ELGs
and ALGs exhibit different properties, they are investigated
separately below. We have assumed that the three ``galaxies'' sources
which remain unclassified (i.e., sources with a ``?'') are ALGs. These
three objects were observed spectroscopically but the signal-to-noise
of their spectra are too low to derive useful spectral types or
redshifts. Given the optical magnitudes of the three sources, any
strong emission lines would have been detected, implying that all
three probably have absorption-line dominated continua.

\subsubsection{Emission-Line Galaxies}\label{ELGs}

We find that 12 out of 18 (67\%) X-ray-detected ELGs have radio
counterparts, while 12 out of 14 (86\%) radio-detected ELGs have X-ray
counterparts. These matching fractions appear to corroborate the
X-ray/IR trend found in Paper~XI. Optically, all of the X-ray/radio
matched ELGs reside within bright, extended galaxies and, when
high-resolution, archival {\it HST} Flanking Field observations are
available, have spiral, irregular, or merging morphologies. As shown
in $\S$\ref{properties}, the radio emission from the majority of the
X-ray/radio matched ELGs sources is indicative of star formation and
yields an average radio spectral slope of $\alpha=0.76\pm0.04$. In
terms of their X-ray emission, however, the spectral constraints for
individual sources are poor. We can, however, place good constraints
on the nature of the X-ray emission from these ELGs in an average
sense. When stacked together, the X-ray/radio detected ELGs have an
average X-ray band ratio\footnote{Defined as the ratio of the
  hard-band count rate to the soft-band count rate.} of $0.20\pm0.04$
(corresponding to $\Gamma=2.02\pm0.09$). This shows that the
fraction of hidden AGN in these sources is low, similar to the
findings of Paper~XI for X-ray/IR ELGs, and that the radio and X-ray
emission are more consistent with active star formation from normal
galaxies \citep[e.g.,][]{Condon1992, Fabbiano1989}. Indeed, the
luminosities of the X-ray/radio matched ELGs span the range expected
from $z\sim0.1$ normal spiral systems up to powerful $z\sim1.3$
starbursts (i.e., $L_{\rm 1.4~GHz}\sim10^{20.7}$--$10^{24.3}$
W~Hz$^{-1}$ and $L_{\rm 0.5-8.0~keV}\sim10^{39.8}$--$10^{42.4}$
\xlum). The ELGs are discussed further in $\S$\ref{discussion}.

\subsubsection{Absorption-Line Galaxies}\label{ALGs}

We find that three of the nine (33\%) X-ray-detected ALGs have radio
counterparts, while three of the four (75\%) radio-detected ALGs have
X-ray counterparts. In contrast to the ELGs, the ALGs have
predominantly elliptical or S0 morphologies. Interestingly, only one
of the X-ray-detected ALGs is clearly an AGN at X-ray energies,
although another two are identified as AGN at radio wavelengths. These
two are the FR~I radio galaxy, CXOHDFN~123644.4$+$621133, and the
variable radio source, CXOHDFN~123652.9$+$621444 (R98). As also found
for the X-ray/radio ELGs, the individual X-ray constraints on these
sources are poor. We can, however, gain some insight into their nature
by stacking their X-ray emission. The average X-ray band ratio of the
three X-ray/radio ALGs is $0.26^{+0.09}_{-0.07}$ (corresponding to
$\Gamma=1.90^{+0.17}_{-0.15}$) and is similar to that for the
X-ray/radio ELGs.\footnote{ Although we note that it is also within
the range typically found for nearby X-ray faint ellipticals and S0s
\citep[e.g.,][]{Irwin1998}.} By comparison, the average band ratio of
the X-ray-detected ALGs that lack radio counterparts is
$0.62^{+0.14}_{-0.12}$ (corresponding to $\Gamma=1.28\pm0.19$),
implying the possibility of different X-ray emission mechanisms at
work. The X-ray luminosities of the three X-ray/radio matched ALGs are
comparable to the most luminous galaxies in the ELG sample (i.e.,
$L_{\rm 0.5-8.0~keV}\sim10^{41.7}$--$10^{42.4}$ \xlum), and they are
generally more luminous than the X-ray-detected ALGs that lack radio
counterparts (i.e., $L_{\rm 0.5-8.0~keV}\sim10^{39.6}$--$10^{41.9}$
\xlum). Such a correspondence between X-ray and radio emission mechanisms 
might be expected from compact low-luminosity AGN
\citep[e.g.][]{Ho1999} or radio jets \citep[e.g.][]{Harris2002}, but
not from stellar processes. Unfortunately, we cannot confirm the
origin of the X-ray emission in these ALGs with the current X-ray
data.

\subsection{Optically Bright AGN}\label{agn}

There are a total of 13 sources that lie between $-1 \la \log (f_{\rm
X}/f_{\rm I}) \la +1$ and are classified as optically bright AGN
($I<24$; OBAGN). Nearly all are confirmed AGN following the criteria
listed in $\S$\ref{properties}. Five of the 13 (38\%) X-ray-detected
OBAGN have radio counterparts, while all three (100\%) radio-detected
OBAGN have X-ray counterparts.

In the nearby Universe, the X-ray spectral properties of luminous AGN
are found to be quite complex and often have a variety of emission and
absorption features \citep[e.g.,][]{Reynolds1997, Turner1997,
George1998, Kaspi2002}. In addition to the commonly seen featureless
continua, X-ray spectra of AGN can also exhibit soft X-ray excesses
\citep[e.g.,][]{Brandt1997, Brandt1999} and absorption via either (1)
``toroidal'' or other material very close to the active nucleus
\citep[e.g., warm absorbers;][]{Reynolds1997, George1998} or (2)
material associated with circumnuclear starbursts
\citep[e.g.,][]{Turner1998, Levenson2001a}.

The wide range of photon indices we find for our higher redshift OBAGN
is qualitatively consistent with the spectral variety seen at low
redshifts (see column~7 of Table~\ref{tab:xray-sources}). 
Interestingly, we find a distinction among the X-ray spectral slopes
of the OBAGN in terms of their radio emission. The average X-ray band
ratio of the radio-detected OBAGN is $0.87\pm0.04$ (corresponding to
$\Gamma=0.95\pm0.04$), while the average band ratio is $0.39\pm0.02$
(corresponding to $\Gamma=1.64\pm0.04$) for radio-undetected OBAGN. 
Even after excluding the X-ray brightest source within the
radio-detected OBAGN sample (CXOHDFN~123646.3$+$621404.6), the average
band ratio is still $0.56\pm0.04$ (corresponding to
$\Gamma=1.36\pm0.07$). If we assume that the OBAGN have intrinsically
similar underlying X-ray continuum shapes, then the above result
implies that the radio-detected OBAGN are more highly obscured at soft
X-ray energies than the radio-undetected OBAGN. To examine the cause
of this spectral difference, we consider the radio and optical
(morphological) properties of the OBAGN below.

X-ray luminous AGN in the local Universe typically have contributions
to their 1.4~GHz radio emission from both star formation processes and
their active nuclei. In fact, the fractions of such objects with
1.4~GHz emission dominated by global star formation and AGN appear to
be roughly comparable \citetext{e.g., \citealp{Wilson1988};
  \citealp{Colbert1996}; J. Gallimore 2002 and J. Ulvelstad 2002,
  priv.\ comm.}. The radio spectral indices and classifications of the
OBAGN in Figure~\ref{fig:r_vs_ar}, coupled with the fact that their
radio luminosity densities are generally modest (i.e., only two OBAGN
have detections or limits above $L_{\rm 1.4~GHz}\sim10^{24}$
W~Hz$^{-1}$), similarly suggest that star formation activity
contributes substantially to the 1.4~GHz emission in at least two of
the five radio-detected OBAGN (CXOHDFN~J123635.6$+$621424,
CXOHDFN~J123642.2$+$621545, and CXOHDFN~J123655.8$+$621201).
Interestingly, the two radio-detected OBAGN with clear starburst-type
radio signatures also appear to have slightly flatter X-ray spectral
slopes than those with AGN-type signatures (compare the values in
Table~\ref{tab:xray-sources}).

Optically, 11 of the 13 OBAGN have been imaged by {\it
HST}.\footnote{The calibrated {\it HST} $I$-band flanking field images
of the HDF-N were retrieved from the {\it HST} archive
(http://archive.stsci.edu/) and were reduced using standard methods.} 
We find that the radio-undetected OBAGN generally have more pronounced
nuclear components (e.g., three are point-like) than the
radio-detected OBAGN and are therefore likely to be less obscured at
optical wavelengths. This fact lends further support to the X-ray
spectral dichotomy above. Even among the radio-detected OBAGN
themselves, we find slight morphological differences. For instance,
the two OBAGN with starburst-type radio spectral slopes have spiral
and irregular morphologies with little or no apparent nuclear
enhancements, while the three OBAGN with AGN-type radio spectral
slopes have elliptical or bulge-dominated spiral morphologies with
moderate nuclear enhancements.

Taken together, the X-ray, optical, and radio properties suggest that
(1) OBAGN with starburst-like optical and radio properties make up a
sizable minority of the population \citetext{perhaps $\sim$20\% of the
OBAGN population; see also \citealp{Heckman1995} and
\citealp{Maiolino1997}} and (2) such sources tend to be more highly
obscured at optical and radio wavelengths compared to other types of
OBAGN. At low redshifts, the star formation activity from such
sources, sometimes called ``X-ray-loud composite galaxies,'' is often
found to be circumnuclear in nature (perhaps as close as $\sim$100~pc
from the central engine) and it is thought to be directly responsible
for the absorption seen at X-ray energies
\citep[e.g.,][]{Levenson2001b}. While we lack the X-ray and radio
spatial resolution needed to address this properly, our findings are
at least consistent with such a scenario.

Finally, starburst-obscured AGN have been postulated to be important
contributors to the cosmic X-ray background (XRB), possibly
outnumbering brighter unobscured AGN by 100 to 1 \citep{Fabian1998}. 
Our findings clearly support the notion that such a population of
AGN exists, even at optical magnitudes of $I<24$. However, the
relative number of starburst-obscured OBAGN is clearly much smaller
than predicted, indicating that either the majority of the presumed
starburst-obscured AGN population lies at fainter optical magnitudes
and presumably higher redshifts (see $\S$\ref{opt_faint}) or that
the overall spectral ``hardening'' of the extragalactic X-ray background (XRB)
is produced by a different mechanism altogether.

\subsection{The Optically Faint X-ray and Radio Sources}\label{opt_faint}

Due to the optical magnitudes of the optically faint ($I\ge24$) X-ray
and radio sources (OFXs and OFRs), their nature remains mostly
uncertain. The best available constraints on these sources to date
suggest that they are comprised of luminous dust-enshrouded starburst
galaxies and luminous obscured AGN at $z\sim1$--$3$, and perhaps a few
extreme redshift ($z>6$) AGN \citetext{\citealp{Richards1999b};
  Paper~VI}. We find that seven of the 23 (30\%) OFXs have radio
counterparts, while seven of the 12 (58\%) OFRs have X-ray
counterparts. Three of the radio-detected OFXs have X-ray spectral
slope constraints that are quite hard (VLA~J123651$+$621221,
VLA~J123707$+$621408, and VLA~J123711$+$621325; see
Table~\ref{tab:xray-sources}) and imply obscured AGN activity, while
the X-ray spectral slope of a fourth radio-detected OFX
(VLA~123655.8$+$621201) is ambiguous. If we stack the three
radio-detected OFXs that lack X-ray spectral constraints
(CXOHDFN~J123636.9$+$621320.3, \break 
CXOHDFN~J123642.1$+$621331.4, and \break
CXOHDFN~J123646.1$+$621448.9), we find a stacked band ratio of
$0.28^{+0.17}_{-0.14}$ ($\Gamma =1.85^{+0.31}_{-0.28}$). This ratio is
significantly different from that found for both the bright
radio-detected OFXs and the OFX population overall ($0.93\pm0.06$, or
equivalently $\Gamma=0.88\pm0.07$; see also Paper~VI), and it suggests
that the origin of the X-ray emission in the faintest radio-detected
OFXs may be due to star formation or Seyfert-luminosity, unobscured
AGN.  Interestingly, the radio spectral indices, classifications, and
luminosity density ranges (assuming $z\sim1$--$3$) for five of the
seven radio-detected OFXs (or 15\% of all OFXs) suggest that
star formation activity is likely to dominate the 1.4~GHz emission.
Thus, to conclude the question of whether starburst-obscured AGN are
important contributors to the XRB (see $\S$\ref{agn}), we find that
the relative number of OFXs that host both obscured AGN and luminous
(i.e. $L_{\rm 1.4~GHz}\ga10^{23}$) starbursts in the CDF-N is also
much smaller than predicted by \citet{Fabian1998}.

\section{Discussion}\label{discussion}

Our analysis of the 1~Ms \hbox{CDF-N} X-ray and 1.4~GHz radio samples
indicates a high degree of overlap between the two source populations
and generally confirms the range of sources types discovered
previously in Papers~II and IV. The highest matching fraction arises
among the X-ray- and radio-detected ELGs and points to a potential
link between the X-ray and radio emission in star-forming galaxies.
The ALG population, on the other hand, appears to contain a variety of
source types, including AGN and normal elliptical/S0 galaxies. Among
the OBAGN population, those which are radio-detected appear to be on
average more obscured than those which are radio-undetected. This
enhanced obscuration may be due to active star formation occurring in
these sources, as implied by their starburst-type radio properties and
optical morphologies. Finally, the radio-detected OFXs appear to
comprise both highly-obscured AGN and unobscured AGN/star-forming
galaxies. A small fraction (i.e., seven out of 36) of the OBAGN and
OFXs display signs of starburst-obscured AGN from their radio
properties.  As such, these sources do not appear to be dominant
contributors to the XRB. We cannot completely rule out the
\citet{Fabian1998} model, however, since several of the remaining
X-ray-detected AGN could still host weak (i.e. $L_{\rm
  1.4~GHz}\la10^{23}$) starbursts.

The X-ray/radio matched ELGs are particularly important from a
cosmological standpoint, as they comprise $\approx$~35\% of the X-ray
source population below X-ray fluxes of $5\times10^{-16}$ \xflux\ and
further mark the emergence of a population of star-forming galaxies
discovered in previous X-ray studies \citetext{e.g.,
Paper~II; Paper~IV; Paper~XI; \citealp{Tozzi2001}}. We further
investigate below the properties of this faint X-ray source
population.

\subsection{An X-ray/radio Correlation?}\label{xray_vs_radio}

Radio emission, because of its tight linear relationship with FIR
emission and its relatively short lifetime due to electron diffusion,
is often used as an estimator of the recent star formation rate
\citep[SFR; e.g.,][]{Condon1992, Yun2001}. To determine whether the X-ray
emission can likewise be used as an indicator of the SFR, we compare
the $K$-corrected full-band X-ray luminosities and 1.4~GHz radio
luminosity densities of the CDF-N ELGs in
Figure~\ref{fig:Lx_vs_Lr}.\footnote{Ideally, we would like to make
comparisons using the hard-band X-ray luminosity in order to minimize
uncertainties related to the unknown and potentially large intrinsic
absorption within the X-ray-detected ELGs. However, very few ELGs have
hard-band detections (22\%). As a compromise, we use instead the
full-band X-ray luminosity since the majority of these sources are
detected in this band (78\%), and the full-band flux is likely to be
less uncertain than the soft-band flux.} The rest-frame X-ray
luminosities of the sources are calculated as

\begin{equation} L_{\rm 0.5-8.0~keV}~=~4~\pi~d_L^2~f_{\rm 0.5-8.0~keV}~(1+z)^{\Gamma-2}~\rm{erg~s^{-1}} \label{eq:lx} \end{equation}

\noindent and the rest-frame radio luminosity densities as

\begin{equation} L_{\rm 1.4~GHz}~=~4~\pi~d_L^2~S_{\rm 1.4~GHz}~10^{-36}~(1+z)^{\alpha-1}~\rm{W~Hz^{-1}}, \label{eq:lr} \end{equation}

\noindent where $d_L$ is the luminosity distance (cm), $f_{\rm X}$ is 
the X-ray flux (erg~cm$^{-2}$~s$^{-1}$), $\Gamma$ corresponds to the
photon index, $S_{\rm 1.4~GHz}$ is the 1.4~GHz flux density ($\mu$Jy),
and $\alpha$ corresponds to the radio spectral index (see
Tables~\ref{tab:xray-sources} and \ref{tab:radio-sources}). We used the
actual measured $\Gamma$ and $\alpha$ values for the $K$-corrections
to the CDF-N ELGs when available, or $\Gamma=2.0$ and $\alpha=0.8$
otherwise. At the redshifts of the sources ($z\sim0.0-1.3$), the
typical $K$-corrections are not large.

Since the majority of the ELGs have upper limits at either X-ray
or radio wavelengths, we need to use survival analysis techniques to
account properly for the many upper limits when searching for a
correlation; we employ the {\sc asurv} statistical package
\citep[v1.2;][]{LaValley1992}, which implements the methods presented
in \citet{Feigelson1985} and \citet{Isobe1986} for censored datasets.

Among the 20 CDF-N ELGs, six have X-ray upper limits and six have
radio upper limits. To establish whether a significant correlation
exists between the X-ray and radio emission, we used the generalized
Kendall's Tau, as it is most applicable for censored datasets with a
small number of points (i.e., $n<30$). The X-ray and radio emission in
ELGs are found to be correlated at an $\approx2.7\sigma$ significance
level. To counter the sparse sampling of the CDF-N ELGs at low
luminosities, we also consider the 102 nearby late-type galaxies
(i.e., non-AGN sources 
\vspace{0.0in}
\centerline{
\includegraphics[width=8.4cm]{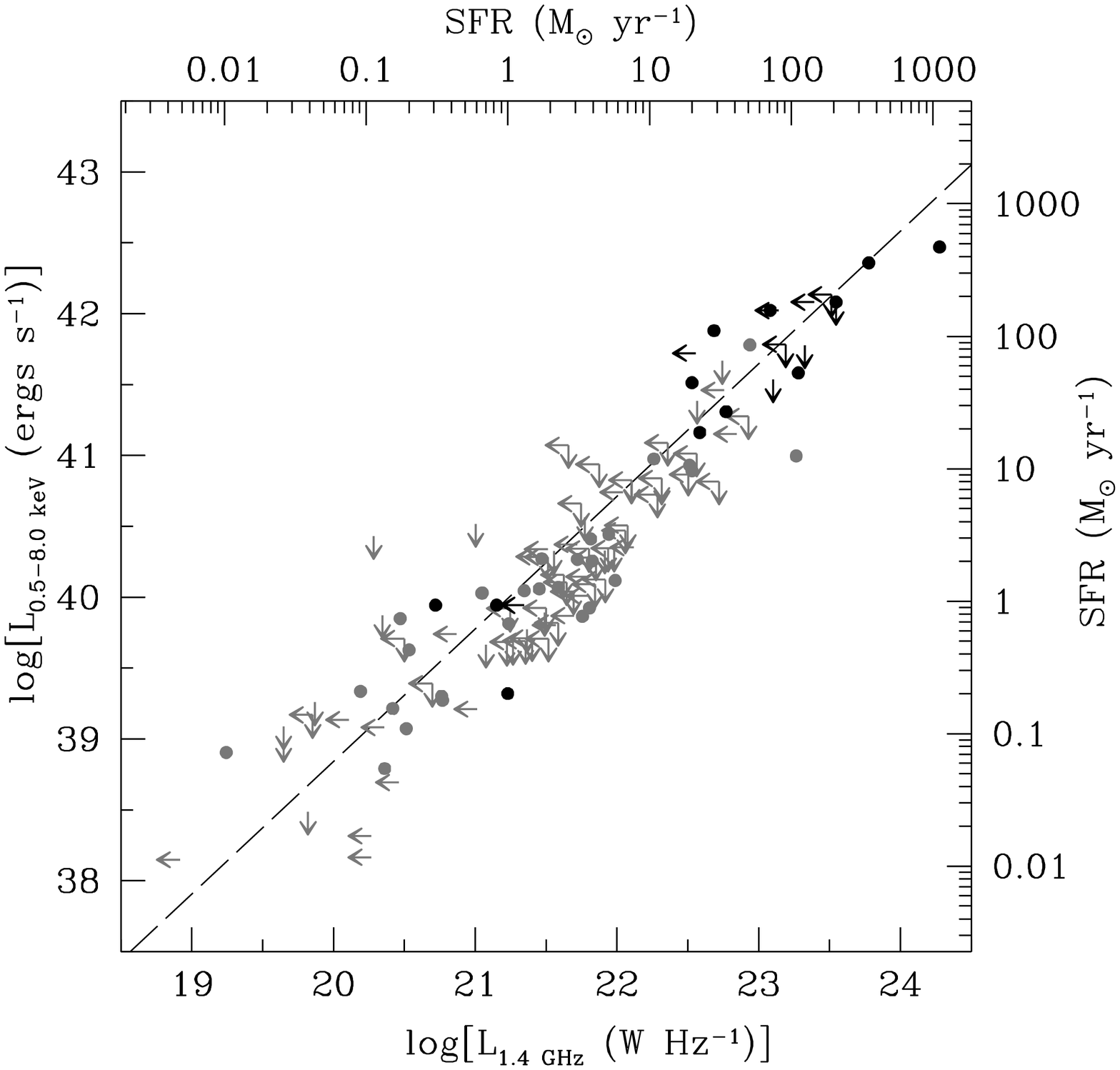}
}
\vspace{-0.1cm} 
\figcaption[fig4.ps]{
A comparison of full-band X-ray luminosity versus 1.4~GHz luminosity
density for the 20 CDF-N X-ray- and radio-detected ELGs (black). Also
plotted are 102 nearby galaxies (gray) from the study of
\citet{Shapley2001}. The luminosities have been $K$-corrected using
the average spectral indices calculated in $\S$\ref{ELGs}. Filled
circles indicate X-ray/radio matched sources, while arrows signify
$3\sigma$ source upper limits. The dashed line shows a Theil-Sen
bisector regression fit to all 122 sources. 
The {\it right} and {\it top} axes indicate the equivalent SFRs
following Equations~\ref{eq:sfr_r}
and \ref{eq:sfr_x}.
\label{fig:Lx_vs_Lr}}
\vspace{0.2in}

\noindent with $L_{\rm X}>10^{38}$~\xlum) from the
multi-wavelength study of \citet[see their Figure 6]{Shapley2001}. The
luminosities of the \citeauthor{Shapley2001} late-type galaxy sample
have been converted to the bands shown in Figure~\ref{fig:Lx_vs_Lr}
assuming spectral slopes of $\Gamma=2.0$ and $\alpha=0.8$,
respectively. Use of the entire 122 galaxy sample dramatically
improves the likelihood of an X-ray/radio correlation to an
$\approx7.7\sigma$ significance level. 

In the presence of upper limits in both variables, the two most
appropriate regression techniques are the Theil-Sen
\citep{Akritas1995} and Schmitt \citep{Schmitt1985} methods. We have
chosen to employ the Theil-Sen bisector regression method since it
does not require (subjective) binning like Schmitt's method, and we
use Schmitt's method only to confirm our
results.\footnote{Instead of defining variables as ``independent'' and
``dependent,'' we obtained regression coefficients (slopes,
intercepts, and the uncertainties in these quantities) for both (X|Y)
and (Y|X), and we then used the bisector of these regressions
following \citet{Isobe1990} and \citet[][Appendix B]{Shapley2001} to
give a more accurate estimate of the relationship between X-ray and
radio emission.}
The Theil-Sen bisector regression analysis for the 122 galaxies yields
the following relationship between the X-ray luminosity and radio
luminosity density:
\begin{equation} \log L_{\rm 0.5-8.0~keV}=(0.935\pm0.073) \log L_{\rm 1.4~GHz} + (20.141\pm1.650).\label{eq:lx_lr}\end{equation}
Schmitt's bisector regression method yields nearly identical results
(i.e., a slope of $0.91\pm0.17$), but it has larger associated
uncertainties.

The X-ray/radio correlation above is consistent with the local
X-ray/radio correlations found both for the \citeauthor{Shapley2001}
sample of late-type galaxies alone and for a sample of nearby
star-forming systems \citep[i.e., sources with optical ``\ion{H}{2}
nuclei'' from the sample of][]{Ho1997} using {\it ASCA} and {\it
BeppoSAX} \citep{Ranalli2002}.
The good agreement between the local correlations and the one
determined above suggests that either (1) there is relatively little
evolution in the X-ray and radio emission of ELGs between $z=0$--$1$,
or (2) the X-ray and radio emission evolve together both in magnitude
and on similar timescales. We further caution that the nearly linear
relationship between X-ray and radio emission found in
Equation~\ref{eq:lx_lr} may not hold for X-ray luminosities below
$\sim10^{39}$~\xlum where the data are sparse. For instance, (1) LMXBs
in globular clusters would have no equivalent long-lasting radio
component and therefore cause the X-ray emission to saturate while the
radio emission continued to decrease, or (2) the relationship between
X-ray emission and SFR may not be linearly proportional for the
population of X-ray binaries emitting at sub-Eddington X-ray luminosities
\citep[e.g.,][]{Grimm2002}.

The physical interpretation of this correlation is complex, as a
variety of emission mechanisms could contribute to the X-ray and radio
emission. In local starburst galaxies, the radio emission is produced
by non-thermal processes and is thought to be driven by supernovae
\citep[although the exact mechanism by which the electrons are accelerated
is still unknown;][]{Condon1992, Bressan2002}. At X-ray energies the
emission is produced by high-mass X-ray binaries
\hbox{(HMXBs; $kT\sim5$--15~keV)} and extended hot gas ($kT\sim0.5$~keV),
which are both attributed to recent star formation, and low-mass X-ray
binaries \hbox{(LMXBs; $kT\sim3$--8~keV),} which trace older star
formation; see, e.g., the review by \citet{Petre1993}. As the majority
of the ELGs lie at $z\sim0.4$--1.3, their observed X-ray fluxes should
be dominated by emission from HMXBs and LMXBs (i.e., emission from
point sources rather than extended hot gas). The majority of the 
\citeauthor{Shapley2001} sample, however, lie below 
$z\sim0.01$, and thus their observed X-ray fluxes may have
contributions from hot gas. Globally, a link between the X-ray and
radio emission seems plausible, given that both are tied to the
evolution of massive stars. However, on an individual HMXB/SNR basis
this link remains quite difficult to test, as these two facets of
massive star formation are temporally distinct. The most significant
progress on the interpretation of this correlation will be from
extensive {\it Chandra} and {\it XMM-Newton} studies of our own Galaxy
and nearby starburst galaxies.

One interesting application of Equation~\ref{eq:lx_lr} is that we can now
use the X-ray emission to determine SFRs for ELGs \citep[e.g.,][]{Grimm2002}. 
Equation~13 of \citet{Yun2001} gives the relationship between the
rest-frame radio luminosity density and SFR, 
\begin{equation} {\rm SFR} (M_{\odot}~{\rm yr}^{-1}) = (5.9 \pm 1.8)\times10^{-22}\  L_{\rm 1.4~GHz} ({\rm W~Hz}^{-1}).\label{eq:sfr_r} \end{equation}
\noindent Substituting Equation~\ref{eq:lx_lr} into the above, we obtain
\begin{equation} {\rm SFR} (M_{\odot}~{\rm yr}^{-1}) = (1.7 \pm 0.5)\times10^{-43}\  L_{\rm 0.5-8.0~keV}^{1.07 \pm 0.08} ({\rm erg~s^{-1}}).\label{eq:sfr_x} \end{equation}

\noindent The implied SFRs found for the X-ray-detected ELGs are
$\sim$~0.2--450~$M_{\odot} {\rm yr}^{-1}$. Similar rates are found
using the associated IR emission from many of the X-ray/radio ELGs
\citep[e.g.,][]{Elbaz2002}, highlighting the consistency across
wavelength bands. Again we caution that Equation~\ref{eq:sfr_x} may
not hold for X-ray luminosities below $\sim10^{39}$~\xlum where the
data are sparse.

\subsection{An {\it ISOCAM} dichotomy?}\label{ISOdichotomy}

In Paper~XI, it was found that 32 of the 40 {\it ISOCAM} 15$\mu$m
sources were identified with ELGs but only $\sim$50\% were 
\vspace{-0.1in}
\centerline{
\includegraphics[width=9.4cm]{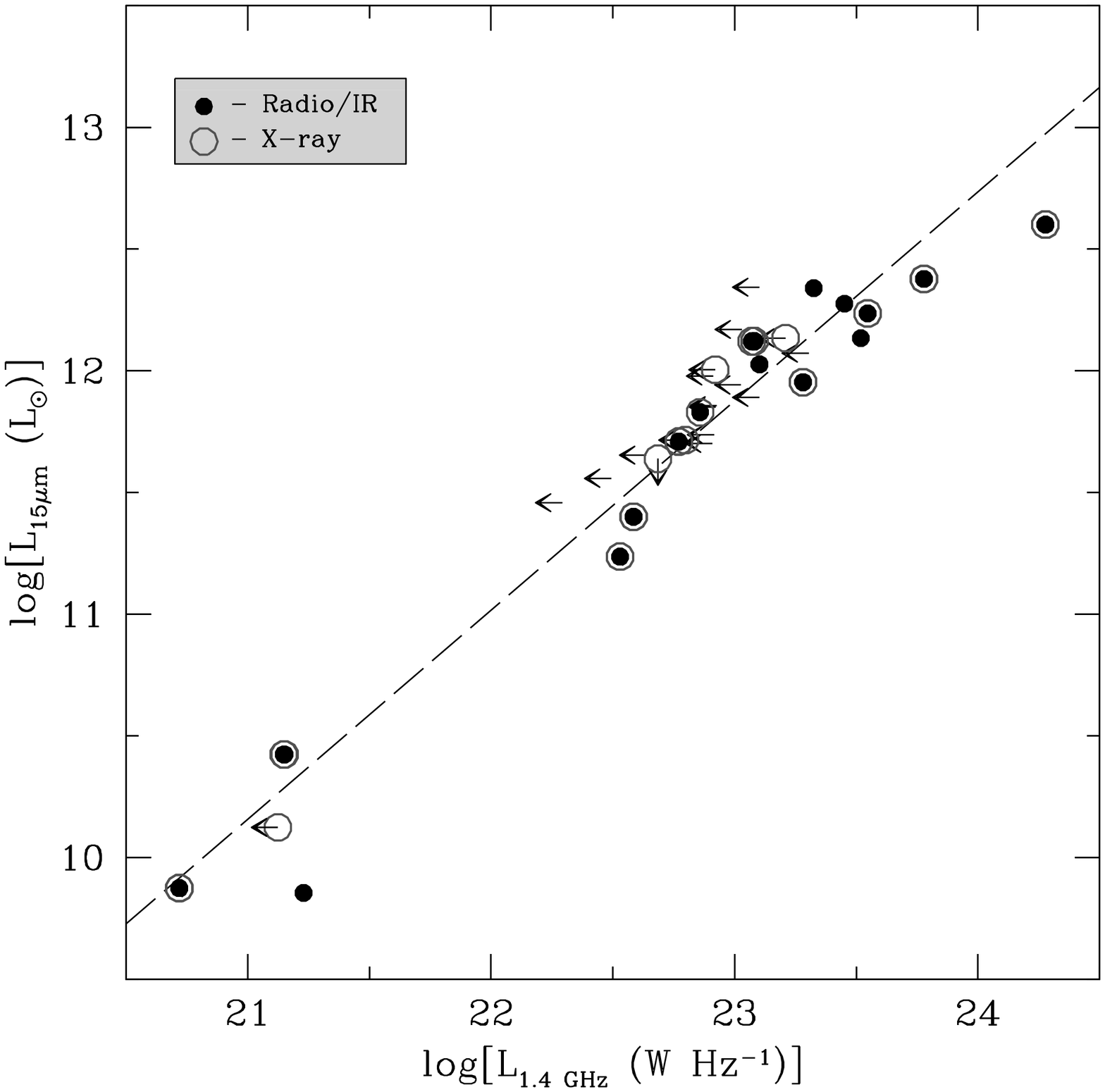}
}
\vspace{0.0cm} 
\figcaption[fig5.ps]{
A comparison of 15$\mu$m luminosity versus 1.4~GHz luminosity density
for all {\it ISOCAM}~15$\mu$m and radio-detected ELGs. Filled circles
indicate 15$\mu$m/radio matched sources, open circles denote X-ray
sources, and arrows signify $3\sigma$ source upper limits. The
dashed line shows the Theil-Sen bisector regression fit to the
sources. \label{fig:15um_vs_r}}
\vspace{0.3cm} 

\noindent
matched with X-ray counterparts. Significantly, the X-ray-detected and
X-ray-undetected 15$\mu$m ELGs were statistically indistinguishable in
terms of their 15$\mu$m fluxes, redshifts, and optical/near-IR
magnitudes, although a much larger fraction of the X-ray detected
15$\mu$m ELGs had radio counterparts. Thus, it appears that the X-ray
and radio emission from ELGs may be better matched than either one is
to the 15$\mu$m emission. This result is somewhat counter to that
expected assuming, e.g., the FIR/radio correlation
\citep[e.g.,][]{Helou1985, Condon1991}, and it suggests that 
a range of source types may exist within the faint {\it ISOCAM}
15$\mu$m source population.

To test whether the 15$\mu$m sources are a homogeneous population,
we compared the 15$\mu$m-detected and radio-detected ELGs using the
complete {\it ISOCAM} 15$\mu$m sample at $>100$~$\mu$Jy of
\citet[][]{Aussel2002} and the radio sample defined in
$\S$\ref{rsample}. The procedure for source matching followed that
outlined in $\S$\ref{sample_match} except that 15$\mu$m sources were
matched to optical sources using a 3$\arcsec$ radius instead of
1$\arcsec$. We find that 12 of the 13 radio-detected ELGs have
15$\mu$m counterparts, with the one remaining radio ELG being
detected by {\it ISOCAM} below the $100$~$\mu$Jy completeness level
\citetext{see source VLA~J123704.6$+$621429 in
Table~\ref{tab:radio-sources} and also \citealp{Aussel1999}}. 
Likewise, 12 of the 32 15$\mu$m-detected ELGs have 1.4~GHz radio
counterparts, with another five ELGs detected as lower significance
1.4~GHz sources.

All of the ELGs from the 15$\mu$m- and radio-detected samples have
redshifts ($z\sim$0.1--1.3), and their resulting $K$-corrected
15$\mu$m luminosities and radio luminosity densities are shown in
Figure~\ref{fig:15um_vs_r}. The rest-frame 15$\mu$m luminosities were
$K$-corrected using an infrared model template of M82
\citep{Silva1998} and the filter efficiencies of {\it ISOCAM} at
15$\mu$m. Note that because we used the SED of M82 to derive the
rest-frame 15$\mu$m luminosities, the 15$\mu$m values will likely be
incorrect for sources with intrinsically different MIR colors. The
radio luminosity densities were calculated following Equation~\ref{eq:lr}. For
each ELG in Figure~\ref{fig:15um_vs_r}, we further indicate whether it
was detected at X-ray energies (this includes both normal and lower
significance detections).

We first note that up to 100\% of the radio-detected ELGs have
15$\mu$m counterparts, suggesting that the FIR/radio relationship
extends to MIR wavelengths and moderate redshifts \citetext{a trait
previously noted by \citealp{Garrett2001} and \citealp{Elbaz2002}}. 
A dashed line denotes the Theil-Sen bisector regression fit to
the 15$\mu$m/radio matched ELGs.

As found in Paper~XI, the majority of the 15$\mu$m-detected ELGs with
radio counterparts also have X-ray counterparts, while most of the
15$\mu$m-detected ELGs without radio counterparts do not. Moreover,
most of the 15$\mu$m-detected ELGs with radio upper limits lie to the
15$\mu$m bright side of the regression. These two facts suggest an
intrinsic bimodality in the 15$\mu$m-detected ELGs.

We can quantify this bimodality statistically by comparing the
15$\mu$m-to-radio luminosity ratio (i.e. reducing the dimensionality
of the problem to single quantity), and then testing for whether the
X-ray-detected distribution differs from the X-ray-undetected one
using the Kaplan-Meier (KM) test within {\sc asurv} to account for the
numerous radio upper limits. The result of the KM test indicates that
the distributions have a $P=0.01$ probability of coming from the same
parent distribution. And since the majority of the X-ray-undetected
15$\mu$m ELGs have radio luminosity density upper limits, this
difference could be even stronger.

The most likely explanation for such a dichotomy is that the
15$\mu$m-detected ELGs with X-ray counterparts have significantly
different MIR spectral energy distributions from those without them. 
To still be consistent with the FIR/radio correlation, these
X-ray-undetected ELGs would have to have substantially warmer MIR/FIR
colors than M82 (i.e., the spectral template used to $K$-correct our
15$\mu$m luminosities). An enhancement to the MIR flux could be
explained by either a hotter dust component or stronger emission-line
features \citep[e.g., PAH features;][]{Puget1989} as compared to M82. 
Because a hotter dust component requires more high energy photons
(i.e., more X-ray emission), however, we might expect enhanced X-ray
emission rather than a lack of it. Thus, the most likely explanation
for the 15$\mu$m-detected ELGs without X-ray counterparts is that
they have significantly stronger emission-line features than those
with X-ray counterparts.

\section{Conclusions}\label{conclusion}

Our main results are the following:

\begin{itemize}
\item We find a large overlap between faint X-ray 
and radio sources detected within $3\arcmin$ of the
\hbox{HDF-N} using the 1~Ms \hbox{CDF-N} and 40~$\mu$Jy VLA
1.4~GHz radio surveys. The matched sources exhibit a broad range of
emission mechanisms, in agreement with the findings of Papers~II and
IV.

\item The highest matching fraction is found among X-ray- and radio-detected
ELGs, which are comprised of apparently normal and starburst galaxies
at redshifts of $z\sim0.1$--1.3 thought to be undergoing recent star
formation. The ELGs are distinct in terms of their X-ray, optical, and
radio properties, and nearly all of them have MIR {\it
ISOCAM} detections.

\item The radio-detected ELGs make up
$\approx$~35\% of the X-ray source population at X-ray fluxes below
$5\times10^{-16}$ \xflux\ and further signal the emergence of a
population of star-forming galaxies discovered in previous X-ray
studies. The high matching fraction among the ELGs suggests that the
deep 40~$\mu$Jy VLA survey at 1.4~GHz is extremely well-matched to the
1~Ms {\it Chandra} observation for the detection of galaxies.

\item The X-ray-emitting AGN population (i.e., OBAGN $+$ OFXs) have
significantly fewer radio matches than the ELGs. These sources exhibit
a range of X-ray spectral slopes suggesting many are moderately
obscured. Moreover, the radio properties and optical morphologies of
the harder X-ray AGN indicate that their obscuration at X-ray energies
may be related to active star formation (perhaps circumnuclear). The
relative numbers of such starburst-obscuring AGN, however, suggest
that they are not likely to be dominant contributors to the XRB.

\item A good correlation exists between X-ray luminosity and radio luminosity 
density for the CDF-N ELGs and nearby late-type galaxies, and it
suggests that the emission mechanisms from the evolution of massive
stars (e.g., the creation of HMXBs) and their eventual destruction
(e.g., supernova-accelerated electrons) are intimately linked, even at
moderate redshifts. This link suggests that X-ray emission can
likewise be used as an indicator of the SFR. The implied SFRs found
for the X-ray-detected ELGs in the CDF-N are
$\sim$~0.2--450~$M_{\odot} {\rm yr}^{-1}$. 

\item Finally, there appear to be two statistically distinct
classes of infrared star-forming galaxies detected with {\it ISOCAM},
one of which shows correlated X-ray and radio emission, and one which
does not. This latter group is likely to have significant stronger
emission-line features than the those with X-ray and radio
counterparts. Confirmation of this result must await the Guaranteed
Time and Great Observatories Origins Deep
Survey\footnote{See http://www.stsci.edu/science/goods/.} observations of
the \hbox{CDF-N} with SIRTF, which will offer both deep spectroscopy
and multi-band photometry at MIR wavelengths over a much larger area. 
We note that the X-ray exposure of the \hbox{CDF-N} has recently
increased to 2~Ms. Thus, the robust X-ray/radio matching fraction we
find with the 1~Ms \hbox{CDF-N} dataset also signals the need for
deeper radio observations with the VLA to build upon this work.

\end{itemize}

\acknowledgements
This work would not have been possible without the support of the
entire {\it Chandra} and ACIS teams; we particularly thank P.~Broos and
L.~Townsley for data analysis software and CTI correction support. 
We thank E.~Feigelson, J. Gallimore, J. H. Schmitt, and J. Ulvestad
for helpful discussions,
M.~LaValley for providing pre-written code for the Theil-Sen estimator
algorithm,
A.~Barger, L.~Cowie, and E.~Richards for kindly providing or making
public their optical and radio images of the \hbox{HDF-N} region.
and the anonymous referee for useful comments that improved the
content and presentation of the paper.
We gratefully acknowledge the financial support of 
NSF CAREER award AST-9983783 (FEB, DMA, WNB, CV), 
NASA GSRP grant NGT5-50247 and the Pennsylvania Space Grant Consortium (AEH), 
NASA grant NAS~8-38252 (GPG, PI), 
and NSF grant AST-9900703~(DPS).

%
%

\def\aa{\tablenotemark{a}}
\def\bb{\tablenotemark{b}}
\def\cc{\tablenotemark{c}}
\def\dd{\tablenotemark{d}}
\def\ee{\tablenotemark{e}}
\def\PP{\tablenotemark{P}}
\def\WW{\tablenotemark{W}}

\begin{deluxetable}{lrrrrrrrrrcl}
\tabletypesize{\scriptsize}
\tablewidth{0pt}
\tablecaption{$3\arcmin$ X-ray Sample\label{tab:xray-sources}} 
\tablehead{
\colhead{(1)} & 
\colhead{(2)} & 
\colhead{(3)} & 
\colhead{(4)} & 
\colhead{(5)} & 
\colhead{(6)} & 
\colhead{(7)} & 
\colhead{(8)} & 
\colhead{(9)} & 
\colhead{(10)} & 
\colhead{(11)} & 
\colhead{(12)} \\ 
\colhead{CXOHDFN Source} & 
\colhead{$I$} & 
\colhead{$S_{\rm 1.4~GHz}$} & 
\colhead{$S_{\rm 8.5~GHz}$} & 
\colhead{$\alpha_{\rm r}$} & 
\colhead{$f_{\rm 0.5-8.0~keV}$} & 
\colhead{$\Gamma$} & 
\colhead{$S_{\rm 15 \mu m}$} &
\colhead{$z$} & 
\colhead{$L_{\rm 0.5-8.0~keV}$} & 
\colhead{Sp. Type} & 
\colhead{Category}
}
\startdata
 123627.3+621257 &    21.94 &      ---     & $<$15.3\aa & \nodata  &  $<$1.4\cc    & \nodata &    --- &      1.221 &     42.23 &   EI & ELG       \\
 123627.3+621308 &    26.14 &      ---     & $<$15.3\aa & \nodata  &     1.1       & \nodata &    --- &    \nodata &   \nodata &  --- & OF        \\
 123627.5+621218 & $>$27.10 &      ---     & $<$16.1\aa & \nodata  &     3.1       &    1.48 &    --- &    \nodata &   \nodata &  --- & OF        \\
 123627.8+621158 &    23.38 &      ---     & $<$17.0\aa & \nodata  &    21.6       & $-$0.17 &    --- &    \nodata &   \nodata &  --- & AGN       \\
 123633.5+621418 &    25.18 &  $<$17.0\aa  & $<$12.0    & \nodata  &    10.0       &    1.74 &    --- &      3.408 &     44.21 &    Q & OF  (AGN) \vspace{0.05in} \\
 123633.7+621313 &    25.79 &  $<$24.9\aa  &  $<$8.7    & \nodata  &    10.6       &$<-$0.29 &    --- &    \nodata &   \nodata &  --- & OF  (AGN) \\
 123633.9+621327 &    23.89 &  $<$25.0\aa  &  $<$8.8    & \nodata  &  $<$1.1\cc    & \nodata &    --- &    \nodata &   \nodata &  --- & ALG?\dd   \\
 123634.4+621213 &    19.21 &    233.0     &    56.5    & 0.74     &     4.3       & $>$1.86 &    448 &      0.458 &     41.88 &    I & ELG       \\
 123634.5+621240 &    22.34 &    230.0     &    52.6    & 0.74     &     3.1       & \nodata &    363 &      1.219 &     42.46 &    E & ELG\dd    \\
 123635.3+621110 &    20.45 &  $<$25.6\aa  & $<$13.6\aa & \nodata  &     9.5       &    0.99 & $<$100 &      0.410 &     41.94 &    A & ALG (AGN) \vspace{0.05in} \\
 123635.3+621152 &    24.30 &  $<$25.0\aa  &  $<$9.5\aa & \nodata  &     2.9       & \nodata &    --- &    \nodata &   \nodata &  --- & OF        \\
 123635.6+621424 &    23.01 &     87.8     &    13.2    & 1.05\bb  &    29.7       &    0.30 &    441 &      2.005 &     43.23 &    E & AGN       \\
 123636.6+621346 &    20.94 &  $<$24.8\aa  &  $<$7.8\aa & \nodata  &   101.0       &    1.82 &    353 &      0.960 &     43.99 &    Q & AGN       \\
 123636.7+621156 &    21.42 &  $<$24.9\aa  &  $<$8.4\aa & \nodata  &    49.4       &    1.74 &  $<$78 &      0.557 &     43.11 &    I & AGN       \\
 123636.9+621320 & $>$26.49 &     50.0     &     7.2\aa & 1.07\bb  &     2.0       & \nodata &    --- &    \nodata &   \nodata &  --- & OF        \vspace{0.05in} \\
 123637.0+621134 &    18.60 &     99.0\WW  &    17.5    & 0.96\bb  &  $<$1.2\cc    & \nodata &    300 &      0.078 &     39.80 &    I & ELG       \\
 123638.5+621339 &    22.31 &  $<$24.6\aa  &  $<$6.8\aa & \nodata  &     5.8       &$<-$0.08 &     52 &      0.357 &     41.18 &    I & AGN       \\
 123639.0+621041 & $>$27.10 &      ---     & $<$15.3\aa & \nodata  &     5.6       &    1.44 &    --- &    \nodata &   \nodata &  --- & OF        \\
 123639.6+621230 &    23.83 &  $<$24.4\aa  &  $<$6.2\aa & \nodata  &     6.9       &    1.87 &  $<$41 &      3.479 &     44.16 &   EQ & AGN       \\
 123639.9+621250 &    21.00 &     36.4\aa  &     9.8    & 0.55\bb  &     2.7       & \nodata &    295 &      0.848 &     42.17 &    I & ELG       \vspace{0.05in} \\
 123640.8+621041 &    24.69 &      ---     & $<$14.2\aa & \nodata  &    21.9       &    0.69 &    --- &    \nodata &   \nodata &  --- & OF (AGN)  \\
 123641.8+621132 &    19.76 &     23.2\aa  &  $<$8.4\aa & $>$0.56\bb &   3.8       & \nodata &    236 &      0.089 &     40.22 &   EI & ELG       \\
 123642.1+621331 &    24.98 &    467.0     &    79.9    & 0.94     &     2.5       & \nodata &     23 &      4.424 &     43.28 &    E & OF  (AGN) \\
 123642.2+621545 &    21.21 &    150.0     &    53.6    & 0.50     &    32.0       &    1.29 &    459 &      0.857 &     43.21 &    I & AGN       \\
 123644.4+621133 &    21.26 &   1290.0     &   599.0    & 0.30     &     2.1       & \nodata &  $<$47 &      1.010 &     42.26 &    A & ALG (AGN) \vspace{0.05in} \\
 123646.1+621448 &    25.21 &    124.0     &    13.3    & 0.84     &     1.4       & \nodata &    --- &    \nodata &   \nodata &  --- & OF        \\
 123646.3+621404 &    21.07 &    179.0     &   190.0    & $-$0.04  &   228.0       &    0.58 &    107 &      0.962 &     43.88 &    Q & AGN       \\
 123646.4+621529 &    21.80 &      ---     & $<$14.8\aa & \nodata  &     5.3       &    1.32 &  $<$92 &    \nodata &   \nodata &  --- & ALG?\dd   \\
 123648.0+621019 &    25.71 &      ---     & $<$16.1\aa & \nodata  &     9.6       &    0.10 &    104 &    \nodata &   \nodata &  --- & OF  (AGN) \\
 123648.1+621308 &    20.29 &  $<$24.0\aa  &  $<$5.5\aa & \nodata  &     7.5       &    1.17 &  $<$22 &      0.476 &     42.01 &    I & ELG       \vspace{0.05in} \\
 123648.3+621426 &    18.60 &     25.1\aa  &     9.8    & 0.57\bb  &     1.5       & \nodata &    307 &      0.139 &     40.21 &    E & ELG       \\
 123648.3+621456 &    26.27 &  $<$25.0\aa  &  $<$9.4\aa & \nodata  &    11.2       &    0.61 &  $<$52 &    \nodata &   \nodata &  --- & OF  (AGN) \\
 123649.4+621346 &    18.00 &  $<$24.2\aa  &  $<$5.6\aa & \nodata  &     0.9       & \nodata &    --- &      0.089 &     39.58 &    A & ALG       \\
 123649.7+621313 &    21.38 &     49.2     &    22.0    & 0.72     &     1.5       & \nodata &    115 &      0.475 &     41.36 &    I & ELG       \\
 123651.1+621031 &    20.17 &     95.0     &    26.0    & 0.74     &     3.0       & \nodata &    341 &      0.410 &     41.52 &    I & ELG       \vspace{0.05in} \\
 123651.3+621051 &    23.55 &      ---     & $<$10.5\aa & \nodata  &    12.7       &$<-$0.11 &    --- &    \nodata &   \nodata &  --- & AGN       \\
 123651.8+621221 &    25.80 &     49.3     &    16.8    & 0.71     &    28.1       &    0.37 &     48 &      2.750 &     43.35 &  --- & OF  (AGN) \\
 123651.8+621505 &    24.65 &      ---     & $<$10.7\aa & \nodata  &    21.3       &    1.43 &    --- &    \nodata &   \nodata &  --- & OF        \\
 123652.9+621444 &    18.92 &    168.0     &   185.0    & $-$0.12  &     5.7       &    1.92 &  $<$53 &      0.322 &     41.67 &    A & ALG (AGN) \\
 123653.4+621139 &    22.05 &     65.7     &    15.1    & 0.77     &     2.2       & \nodata &    180 &      1.275 &     42.45 &   EA & ELG       \vspace{0.05in} \\
 123653.6+621115 &    22.47 &  $<$24.9\aa  &  $<$8.3\aa & \nodata  &  $<$1.4\cc    & \nodata &  $<$80 &    \nodata &   \nodata &  --- & ELG?      \\
 123654.6+621111 &    26.07 &  $<$25.0\aa  &  $<$8.8\aa & \nodata  &     7.7       &    0.28 &    --- &    \nodata &   \nodata &  --- & OF  (AGN) \\
 123655.4+621311 &    21.92 &  $<$23.6\aa  &    12.3    & $<$0.36  &    12.7       &    1.50 &     23 &      0.968 &     43.00 &    A & AGN       \\
 123655.8+621201 &    24.13 &     24.3\aa  &  $<$6.7\aa & $>$0.71\bb &   4.3       &    1.36 &    --- &   1.140\PP &     42.54 &  --- & OF (AGN?)      \\
 123656.6+621245 &    20.24 &  $<$24.2\aa  &  $<$5.6\aa & \nodata  &     1.1       & \nodata &  $<$28 &      0.518 &     41.32 &    A & ALG       \vspace{0.05in} \\
 123656.6+621513 &    26.20 &      ---     & $<$13.2\aa & \nodata  &     6.0       &$<-$0.17 &    --- &    \nodata &   \nodata &  --- & OF  (AGN) \\
 123656.9+621301 &    22.66 &     49.5     &     7.6    & 1.04\bb  &     8.6       &    1.77 &  $<$45 &      0.474 &     42.20 &    E & ELG\dd    \\
 123657.3+621024 &    22.35 &      ---     & $<$17.1\aa & \nodata  &     2.7       & \nodata &    --- &      0.847 &     41.86 &    E & ELG       \\
 123657.5+621210 &    20.71 &  $<$24.4\aa  &  $<$6.2\aa & \nodata  &     2.4       & \nodata &  $<$80 &      0.665 &     41.89 &    A & ALG       \\
 123658.8+621022 &    24.28 &      ---     & $<$18.5\aa & \nodata  &    12.0       &$<-$0.34 &    --- &    \nodata &   \nodata &  --- & OF  (AGN) \vspace{0.05in} \\
 123658.8+621435 &    20.40 &  $<$24.4     &    11.4    & $<$0.42  &    73.1       &    1.48 &  $<$50 &      0.678 &     43.41 &    I & AGN       \\
 123700.4+621509 &    23.89 &      ---     & $<$14.7\aa & \nodata  &    16.6       &    1.11 &    --- &    \nodata &   \nodata &  --- & AGN       \\
 123702.0+621123 &    18.53 &  $<$25.3\aa  & $<$10.9\aa & \nodata  &     1.5       & \nodata &    162 &      0.136 &     40.21 &    I & ELG       \\
 123702.6+621244 &    23.97 &  $<$24.7\aa  &  $<$7.2\aa & \nodata  &     3.5       & \nodata &    --- &    \nodata &   \nodata &  --- & AGN       \\
 123704.1+621155 &    25.66 &  $<$25.1\aa  &  $<$9.7\aa & \nodata  &     1.6       & \nodata &    --- &    \nodata &   \nodata &  --- & OF        \vspace{0.05in} \\
 123704.6+621429 &    20.78 &     25.7\aa  & $<$12.5\aa & $>$0.37\bb &   2.3       & \nodata &     72 &      0.561 &     41.70 &    I & ELG       \\
 123707.2+621408 &    24.64 &     45.3     &    29.0    & 0.29     &    10.0       &    0.50 &    --- &    \nodata &   \nodata &  --- & OF  (AGN) \\
 123711.4+621331 &    22.96 &    132.0     &    31.1    & 0.69     &     2.9       & \nodata &    --- &   1.110\PP &     42.37 &  --- & ALG?\dd   \\
 123712.0+621325 &    24.87 &     53.9     &    13.5\aa & 0.76\bb  &     5.6       & $<$0.09 &    --- &    \nodata &   \nodata &  --- & OF  (AGN) \\
 123712.1+621211 &    25.90 &      ---     & $<$17.7\aa & \nodata  &     1.9       & \nodata &    --- &    \nodata &   \nodata &  --- & OF        \vspace{0.05in} \\
 123712.7+621342 & $>$26.20 &      ---     & $<$18.6\aa & \nodata  &     1.6       & \nodata &    --- &    \nodata &   \nodata &  --- & OF        \\
 123714.4+621221 &    23.05 &      ---     & $<$21.3\aa & \nodata  &     1.7       & \nodata &    --- &      1.084 &     42.19 &    E & ELG\dd    \\
\enddata
\tablecomments{
{\it Col.~1}: Source name given as CXOHDFN~JHHMMSS.S$+$DDMMSS. 
{\it Col.~2}: $I$-band magnitude. See $\S$\ref{sample_match} for details.
{\it Cols.~3 \& 4}: 1.4~GHz and 8.5~GHz catalog radio flux densities
in $\mu$Jy, respectively. An ``a'' indicates that the 1.4~GHz or
8.5~GHz flux densities have been measured using the publicly available
images down to $3\sigma$ limit. A `---' indicates that the source lies
outside of the publicly available VLA 1.4~GHz image and thus the
source has no detection down to a $5\sigma$ flux density limit of
$\sim$40~$\mu$Jy at 1.4~GHz. The 1.4~GHz flux density for the source
denoted by a ``W'' is taken from the the WSRT 1.4~GHz catalog of
\citet{Garrett2000}.
{\it Col.~5}: Radio spectral slope, $\alpha_{\rm r}$, calculated using
1.4~GHz and 8.5~GHz fluxes measured in their respective 3\farcs5
convolved images (R00). A ``b'' indicates that the spectral slopes
have been calculated using lower significance 1.4~GHz and/or 8.5~GHz
flux densities measured from the publicly available 1\farcs9 and
3\farcs5 convolved images, respectively. As such, these spectral
slopes should therefore only be considered approximate since the
beamsizes of the 1.4~GHz and 8.5~GHz images have not been properly
matched.
{\it Col.~6}: Observed 0.5--8.0~keV flux calculated assuming the value
of $\Gamma$ listed in Col.~7 or $\Gamma=1.4$ if not listed (see
Paper~V for details), in units of 10$^{-16}$~erg~cm$^{-2}$~s$^{-1}$. 
Sources denoted by a ``c'' are detected only in the 0.5--2.0~keV band. 
We calculated X-ray upper limits following $\S$3.2.1 of Paper~V.
{\it Col.~7}: X-ray power-law photon index, $\Gamma$, computed from
the band ratio (i.e., the ratio of 2.0--8.0~keV count rate, or
3$\sigma$ upper limit, to 0.5--2.0~keV count rate; see Paper~V for
details). Note that many sources lack $\Gamma$ values due to their
low number of counts.
{\it Col.~8}: {\it ISOCAM} 15$\mu$m flux density as reported in
\citet{Aussel1999, Aussel2002}. Three $\sigma$ 15$\mu$m
upper limits were computed for all X-ray sources within the complete {\it
ISOCAM} area following \citet{Aussel2002}. A `---'
indicates that the source lies outside of the {\it ISOCAM} field of
view.
{\it Col.~9}: Spectroscopic redshift from the catalogs of
\citet{Cohen2000} or \citet{Dawson2001}. The two redshifts denoted by
``P's'' are photometric and were obtained from the catalog presented in
Paper~IX.
{\it Col.~10}: Logarithm of the absorption-corrected rest-frame
0.5--8.0~keV luminosities, calculated assuming the
unabsorbed spectrum is a $\Gamma=2.0$ power law, in units of \xlum.
{\it Col.~11}: Spectral classification from the catalogs of
\citet{Cohen2000}, or if from \citet{Dawson2001} or \citet{Barger2002}, 
converted to the \citet{Cohen2000} classification scheme.
{\it Col.~12}: Source classification category (see
$\S$\ref{properties}). An ``d'' indicates that the source lies in the
AGN region of Figure~\ref{fig:x_vs_i-radio} but is considered to be
part of the galaxy sample. Entries appended with a ``?'' are
tentative.}
\end{deluxetable}

%
%

\begin{deluxetable}{lrrrrrrrrrccl}
\tabletypesize{\scriptsize}
\tablewidth{0pt}
\tablecaption{$3\arcmin$ 1.4~GHz Sample\label{tab:radio-sources}} 
\tablehead{
\colhead{(1)} & 
\colhead{(2)} & 
\colhead{(3)} & 
\colhead{(4)} & 
\colhead{(5)} & 
\colhead{(6)} & 
\colhead{(7)} & 
\colhead{(8)} & 
\colhead{(9)} & 
\colhead{(10)} & 
\colhead{(11)} & 
\colhead{(12)} &
\colhead{(13)} \\
\colhead{VLA Source} & 
\colhead{$I$} & 
\colhead{$S_{\rm 1.4~GHz}$} & 
\colhead{$S_{\rm 8.5~GHz}$} & 
\colhead{$\alpha_{\rm r}$} & 
\colhead{$f_{\rm 0.5-8.0~keV}$} & 
\colhead{$\Gamma$} & 
\colhead{$S_{\rm 15 \mu m}$} &
\colhead{$z$} & 
\colhead{$L_{\rm 0.5-8.0~keV}$} & 
\colhead{Sp. Type} & 
\colhead{Radio Type} & 
\colhead{Category} 
}
\startdata
 123634.4+621212 &    19.21 &   233.0     &    56.5    &   0.74    &    4.3       & $>$1.86 &    448 &    0.458 &       41.88 &    I & S & ELG       \\
 123634.5+621241 &    22.34 &   230.0     &    52.6    &   0.74    &    3.1       & \nodata &    363 &    1.219 &       42.47 &    E & S & ELG\dd    \\
 123635.6+621424 &    23.01 &    87.8     &    13.2\aa &   1.05\bb &   29.7       &    0.30 &    441 &    2.005 &       43.23 &    E & S & AGN       \\
 123636.9+621320 & $>$26.50 &    50.0     &     7.2\aa &   1.07\bb &    2.0       & \nodata &    --- &  \nodata &     \nodata &  --- & U & OF        \\
 123640.7+621010 &    25.84 &    86.8     &    29.2    &   0.44    & $<$1.5       & \nodata &    --- &  \nodata &     \nodata &  --- & U & OF        \vspace{0.05in} \\
 123642.0+621331 &    24.98 &   467.0     &    79.9    &   0.94    &    2.5       & \nodata &     23 &    4.424 &       43.28 &    E & U & OF  (AGN) \\
 123642.2+621545 &    21.21 &   150.0     &    53.6    &   0.50    &   32.0       &    1.29 &    459 &    0.857 &       43.21 &    I & S & AGN       \\
 123644.3+621133 &    21.26 &  1290.0     &   599.0    &   0.30    &    2.1       & \nodata &  $<$41 &    1.010 &       42.26 &    A & A & ALG (AGN) \\
 123646.0+621448 &    25.21 &   124.0     &    13.3    &   0.84    &    1.4       & \nodata &    --- &  \nodata &     \nodata &  --- & U & OF        \\
 123646.3+621404 &    21.07 &   179.0     &   190.0    &$-$0.04    &  228.0       &    0.58 &    107 &    0.962 &       43.88 &    Q & A & AGN       \vspace{0.05in} \\
 123646.6+621226 & $>$26.20 &    72.0     &    15.6\aa &   0.85\bb & $<$0.7       & \nodata &$<$100 &  \nodata &     \nodata &  --- & U & OF        \\
 123646.7+621445 &    23.22 &   117.0     &  $<$9.1\aa &$>$1.42\bb & $<$1.0       & \nodata &    --- &  \nodata &     \nodata &  --- & S & ALG       \\
 123649.7+621312 &    21.38 &    49.2     &    22.0    &   0.72    &    1.5       & \nodata &    115 &    0.475 &       41.36 &    I & S & ELG       \\
 123651.1+621030 &    20.17 &    95.0     &    26.0    &   0.74    &    3.0       & \nodata &    341 &    0.410 &       41.52 &    I & S & ELG       \\
 123651.7+621221 &    25.80 &    49.3     &    16.8    &   0.71    &   28.1       &    0.37 &     48 &    2.750 &       43.35 &  --- & S & OF  (AGN) \vspace{0.05in} \\
 123652.9+621444 &    18.92 &   168.0     &   185.0    &$-$0.12    &    5.7       &    1.92 &  $<$53 &    0.322 &       41.67 &    A & A & ALG (AGN) \\
 123653.4+621139 &    21.96 &    65.7     &    15.1    &   0.77    &    2.2       & \nodata &    180 &    1.275 &       42.45 &   EA & S & ELG       \\
 123654.7+621039 & $>$26.20 &    48.2     & $<$13.7\aa &$>$1.00    & $<$0.6       & \nodata &    --- &  \nodata &     \nodata &  --- & U & OF        \\
 123656.6+621207 & $>$26.20 &    46.2     &  $<$6.7\aa &$>$1.07    & $<$0.9       & \nodata &    --- &  \nodata &     \nodata &  --- & U & OF        \\
 123656.9+621302 &    22.66 &    49.5     &     7.6\aa &   1.04\bb &    8.6       &    1.77 &    --- &    0.474 &       42.20 &    E & S & ELG\dd    \vspace{0.05in} \\
 123659.9+621449 &    21.20 &    47.0     & $<$12.1\aa &$>$0.75\bb & $<$1.1       & \nodata &    295 &    0.762 &    $<$41.46 &   EI & U & ELG       \\
 123701.5+621146 &    25.30 &   128.0     &    29.5    &   0.67    &    1.4       & \nodata &$<$100  &    0.884 &       41.70 &    E & S & OF        \\
 123702.8+621401 &    22.17 &    41.4     &  $<$9.5\aa &$>$0.79    & $<$1.2\cc    & \nodata &    144 &    1.243 &    $<$41.98 &    E & S & ELG       \\
 123705.8+621153 &    20.60 &    52.5     & $<$11.9\aa &$>$0.82\bb & $<$1.3       & \nodata &    431 &    0.904 &    $<$41.69 &    I & S & ELG       \\
 123707.2+621408 &    24.64 &    45.3     &    29.0    &   0.29    &   10.0       &    0.50 &    --- &  \nodata &     \nodata &  --- & U & OF  (AGN) \vspace{0.05in} \\
 123707.9+621121 &    25.30 &    60.3     &     6.2\aa &   1.26\bb & $<$1.4       & \nodata &    --- &  \nodata &     \nodata &  --- & U & OF        \\
 123711.3+621331 &    22.96 &   132.0     &    31.1    &   0.69    &    2.9       & \nodata &    --- & 1.110\PP &       42.37 &  --- & S & ALG?      \\
 123711.9+621325 &    24.87 &    53.9     &    13.5\aa &   0.76\bb &    5.6       & $<$0.09 &    --- &  \nodata &     \nodata &  --- & S & OF  (AGN) \\
\enddata
\tablecomments{
{\it Col.~1}: Source name given as VLA~JHHMMSS.S$+$DDMMSS. 
{\it Col.~2}: $I$-band magnitude. See $\S$\ref{sample_match} for details.
{\it Cols.~3 \& 4}: 1.4~GHz and 8.5~GHz catalog radio flux densities
in $\mu$Jy, respectively. An ``a'' indicates that the 8.5~GHz flux
densities have been measured using the publicly available image down to a 
$3\sigma$ limit.
{\it Col.~5}: Radio spectral slope, $\alpha_{\rm r}$, calculated using
the 1.4~GHz and 8.5~GHz fluxes measured in their respective 3\farcs5
convolved images (R00). A ``b'' indicates that the spectral slopes
have been calculated using lower significance 8.5~GHz flux densities
measured from the publicly available 3\farcs5 convolved image. As
such, these spectral slopes should therefore only be considered
approximate since the beamsizes of the 1.4~GHz and 8.5~GHz images have
not been properly matched.
{\it Col.~6}: Observed 0.5--8.0~keV flux calculated assuming the value
of $\Gamma$ listed in Col.~7 or $\Gamma=1.4$ if not listed (see
Paper~V for details), in units of 10$^{-16}$~erg~cm$^{-2}$~s$^{-1}$. 
Sources denoted by a ``c'' are detected only in the 0.5--2.0~keV band. 
We calculated X-ray upper limits following $\S$3.2.1 of Paper~V.
{\it Col.~7}: X-ray power-law photon index, $\Gamma$, computed from
the band ratio (i.e., the ratio of 2.0--8.0~keV count rate, or
3$\sigma$ upper limit, to 0.5--2.0~keV count rate; see Paper~V for
details). Note that many sources lack $\Gamma$ values due to 
their low number of counts.
{\it Col.~8}: {\it ISOCAM} 15$\mu$m flux density as reported in
\citet{Aussel1999, Aussel2002}. Three $\sigma$ 15$\mu$m
upper limits were computed for all X-ray sources within the complete {\it
ISOCAM} area following \citet{Aussel2002}. A `---'
indicates that the source lies outside of the {\it ISOCAM} field of
view.
{\it Col.~9}: Redshift from the catalogs of \citet{Cohen2000} or
\citet{Dawson2001}. The one redshift denoted by a ``P'' is photometric and
was obtained from the catalog presented in Paper~IX.
{\it Col.~10}: Logarithm of the absorption-corrected rest-frame
0.5--8.0~keV luminosities, calculated assuming the
unabsorbed spectrum is a $\Gamma=2.0$ power law, in units of \xlum.
{\it Col.~11}: Spectral classification from the catalogs of
\citet{Cohen2000}, or if from \citet{Dawson2001} or
\citet{Barger2002}, converted to the \citet{Cohen2000} classification
scheme.
{\it Col.~12}: Radio classification from R99.
{\it Col.~13}: Source classification category (see
$\S$\ref{properties}). An ``d'' indicates that the source lies in the
AGN region of Figure~\ref{fig:x_vs_i-radio} but is considered to be
part of the galaxy sample. Entries appended with a ``?'' are
tentative.}
\end{deluxetable}

\end{document}